%% file: fvdets_3.tex
\documentclass[reqno]{amsproc}
\usepackage{amssymb}
\usepackage{amsthm}
\usepackage{physics}
\usepackage{mathtools}
\usepackage{tikz}
\usetikzlibrary{arrows.meta}
\tikzset{> = {Straight Barb[scale=.75]}}

\usepackage[citation-order,nobysame]{amsrefs}
\usepackage{xyzbib}

\numberwithin{equation}{section}
\mathtoolsset{showonlyrefs}

\let\cite=\cites

\newtheorem{theorem}{Theorem}
\newtheorem{proposition}[theorem]{Proposition}
\newtheorem{lemma}[theorem]{Lemma}

\DeclareMathOperator\sgn{\mathrm{sgn}}

\newcommand{\pd}{\partial}
\newcommand{\ua}{\uparrow}

\newcommand{\ii}{\mathrm{i}}

\newcommand{\hatn}{\hat N}

\newcommand{\Ftwoone}[4]{%
\,{}_{2}F_{1}\bigg(\genfrac{}{}{0pt}{}{#1,\, #2}{#3} \bigg\vert #4\bigg)}

\begin{document}

\title{Determinant formulas for the five-vertex model}
\author{Ivan N. Burenev}
\address{Steklov Mathematical Institute, 
Fontanka 27, St.~Petersburg, 191023, Russia}
\email{inburenev@gmail.com}

\author{Andrei G. Pronko}
\address{Steklov Mathematical Institute, 
Fontanka 27, St.~Petersburg, 191023, Russia}
\email{agp@pdmi.ras.ru}

\begin{abstract}
We consider the five-vertex model on a finite square lattice with 
fixed boundary conditions such that the configurations of the model 
are in a one-to-one correspondence with the boxed plane partitions 
(3D Young diagrams which fit into a box of given size). 
The partition function of an inhomogeneous model is given 
in terms of a determinant. For the homogeneous model, it 
can be given in terms of a Hankel determinant. We also show that
in the homogeneous case the partition function 
is a $\tau$-function of the sixth Painlev\'e equation with respect
to the rapidity variable of the weights.  
\end{abstract}
\maketitle
\tableofcontents

\section{Introduction}

In studying vertex models on finite lattices an important role is played 
by representations in terms of determinants for their partition functions. 
A famous example is provided by the six-vertex model with domain wall 
boundary conditions \cite{I-87} as well as with their modifications 
related to symmetry classes of alternating-sign matrices \cite{Ku-02}.  

The five-vertex model is the six-vertex model with one of its six
(actually one of its first four, numbered in the standard order \cite{B-82,LW-72}) 
vertices frozen out. Historically, the five-vertex model 
was first considered in the context of modeling of a crystal growth or evaporation 
in two dimensions, based on the terrace-ledge-kink picture of a crystal surface 
\cite{G-90,GLT-90}. In a different crystal growth picture, 
it also admits an interpretation as a probabilistic cellar automaton \cite{GS-92}.   
Recent interest to the five-vertex model is motivated, in particular, 
by its close connection with Grothendieck polynomials 
\cite{MSa-13,MSa-14,M-20,BBBG-21,BS-20,GZj-20}. It is also known, that
in the case of a finite lattice with special fixed boundary conditions
the five-vertex model configurations are in one-to-one correspondence with 
boxed plane partitions (3D Young diagrams which fit into a box of given size)
\cite{B-10,P-16}. 

In the present paper, we prove determinant representations for the partition function   
of the five-vertex model, which arise in the latter context. 
In \cite{B-10}, it was shown that in the case where the configurations of the model 
are in correspondence with boxed plane partitions which fit into a box with 
two equal sides, the partition function can be written 
as a scalar product of off-shell Bethe 
states and it evaluates in terms of a determinant. Our aim here is 
to extend this result to the case where the configurations of the model 
are in correspondence with boxed plane partitions which fit into 
a box with no restriction on the sides of the box.
In the algebraic Bethe Ansatz, the partition function in this case 
can be written as the matrix element of a product of $A$- or $D$-operators 
between off-shell Bethe states. For these matrix elements we prove 
formulas which express them in terms of determinants of the sizes equal to 
the total number of operators between the vacuum states.

\subsection{The model}

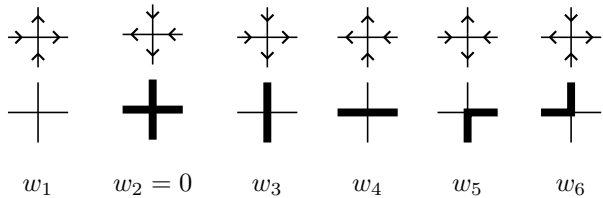
\begin{figure}
\centering
\input{fig-SixVertices}
\caption{The six vertices of the six-vertex model in terms of 
arrows (first row) or lines (second row), and their Boltzmann weights 
in the five-vertex model (third row)}
\label{fig-SixVertices}
\end{figure}

The five-vertex model is a special case of the six-vertex
model. The configurations of the model can be
represented in terms of 
arrows placed on edges of a square lattice or, equivalently, in terms of
lines  ``flowing'' through the lattice. We use the convention 
\cite{LW-72,B-82} between the arrow and line pictures of the configurations that 
if an arrow points down or left, then
this edge has a line, otherwise the edge is empty. 
In the six-vertex model the admissible vertices are only those which have 
equal number of incoming and outgoing arrows, see Fig.~\ref{fig-SixVertices} where 
the vertices are shown in the standard order. 
The five-vertex model can be obtained by  
requiring that only those vertices are admitted which contain non-intersecting 
lines, that is, the second type of the vertex is forbidden. 

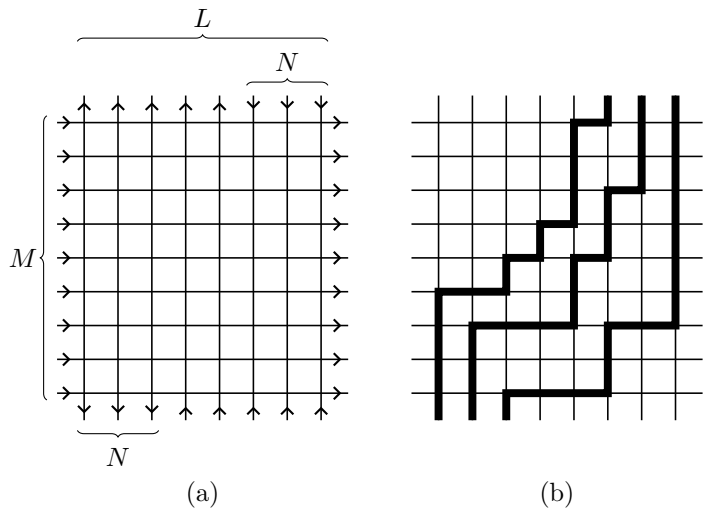
\begin{figure} 	
\centering
\input{fig-LxMlattice}
\caption{An $L\times M$ lattice with $N$ first (last) down arrows 
at the bottom (top) boundary and the remaining arrows 
pointing up or right ($L=8$, $M=9$, $N=3$): 
(a) Boundary conditions in terms of arrows, (b) An example 
of configuration of the five-vertex model in terms of lines}
\label{fig-LxMlattice}
\end{figure}

\begin{figure} 	
\centering
\input{fig-3DYoung}
\caption{The configuration of the five-vertex model shown in 
Fig.~\ref{fig-LxMlattice}b as a projection on the plane of a 
3D Young diagram which fits into $A\times B\times C$ box, 
where $A=L-N$, $B=N$, and $C=M-N$ (left), and 
the correspondence between vertices and flat fragments 
of images of 3D Young diagrams (right)}
\label{fig-3DYoung}
\end{figure}
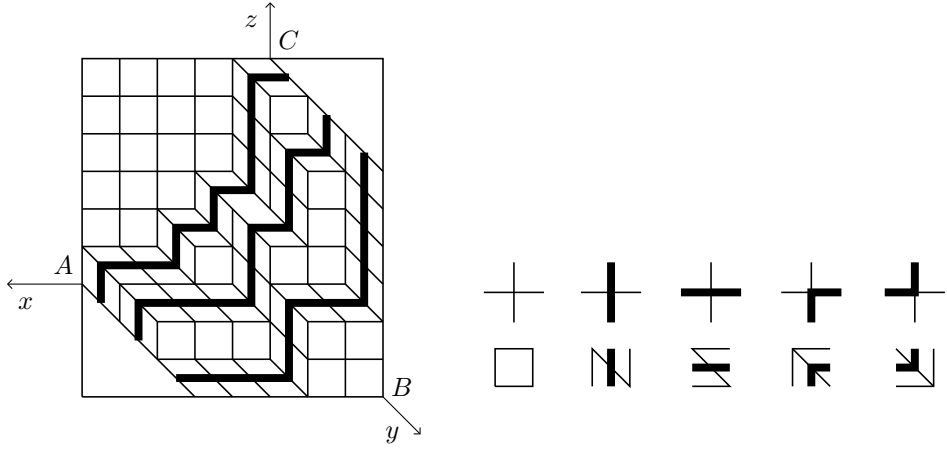

In this paper we consider the model on a lattice obtained by intersection of 
$L$ vertical and $M$ horizontal lines (the $L\times M$ lattice). We are interested 
in the special fixed boundary conditions, such that $N$ first (last) arrows 
at the bottom (top) boundary point down, and the remaining arrows
point up or right, see Fig.~\ref{fig-LxMlattice}.      
With these boundary conditions, the configurations of the model are in one-to-one 
correspondence with plane partitions, or 
3D Young diagram, which fits into $A\times B\times C$ box, 
where $A=L-N$, $B=N$, and $C=M-N$, see Fig.~\ref{fig-3DYoung}. 
In this correspondence, the lines of the vertex model are gradient lines;
there also exists the one-to-one correspondence between vertices 
and flat fragments of images of 3D Young diagrams.

To define the partition function, we introduce the following 
parameterization of the Boltzmann weights $w_i=w_i(u)$
in terms of a spectral parameter $u$:
\begin{equation}\label{weights}
w_1=\frac{\alpha}{\Delta}\left(u-\frac{1}{u}\right),\qquad
w_3=\frac{u}{\alpha},\qquad
w_4=\alpha u,\qquad
w_5=w_6=1.
\end{equation}
Here, the parameter $\alpha$ plays the role of an external field, and
$\Delta$ is the parameter which describes an interaction, similarly to the 
parameter $\Delta$ in the six-vertex model (see, e.g., \cite{LW-72,B-82}); 
note that for the five-vertex model $\Delta=(w_3w_4-w_5w_6)/w_1w_3$. 

Given parameterization \eqref{weights}, associate to each vertical line
a parameter $u_j$, $j=1,\ldots,L$, and to each 
horizontal line a parameter $\xi_k$, $k=1,\ldots,M$. We count lines from 
right to left and from top to bottom. The partition 
function is defined as follows:
\begin{equation}\label{ZLM}
Z=\sum_{\mathcal{C}}
\prod_{j=1}^{L}
\prod_{k=1}^{M}W_{j,k}(\mathcal{C}).
\end{equation}
Here, the sum is performed over all possible arrow 
configurations $\mathcal{C}$ of the model on the $L\times M$ lattice.  
The function $W_{j,k}(\mathcal{C})$ is the 
Boltzmann weight of the vertex at the intersection of the 
$j$th vertical and $k$th horizontal lines, and 
$W_{j,k}(\mathcal{C})=w_i(u_j/\xi_k)$, where the value of $i=1,3,4,5,6$
is determined by the configuration $\mathcal{C}$.  

By the construction, the partition function is a function of the 
two sets of variables $u_1,\ldots,u_L$ and $\xi_1,\ldots,\xi_M$, where 
the variables are assumed not to be equal to each other within each set. 
From the statistical mechanics point of view, it 
is also interesting to consider the case where these variables are equal to 
each other within each set, that is 
$u_1=\dots=u_L=:u$ and $\xi_1=\dots=\xi_M=:\xi$, where, without loss of generality, 
one can set $\xi=1$. Then the partition 
function is just a function of the variable $u$ (and of 
the parameters $\alpha$ and $\Delta$). We refer 
to the limit $u_1,\ldots,u_L\to u$ and $\xi_1,\ldots,\xi_M\to 1$
as \emph{homogeneous limit}.      

\subsection{Main results}

The main results of the paper concern determinant formulas for the 
partition function. Namely, we state and prove two results, one 
is about the partition function \eqref{ZLM} and second is about
its value in the homogeneous limit. 

The first result can be formulated as follows.
 
\begin{theorem}\label{Th1}
For the partition function of the inhomogeneous five-vertex model 
on the $L\times M$ lattice the following representation is valid:
\begin{equation}\label{ZdetVLN}
Z=\Delta^{(L-N)N} \prod_{j=1}^{|L-2N|}u_j^{\sgn(L-2N)}
\prod_{1\leq i<j\leq L}^{} \frac{1}{u_j^2-u_i^2} \det \mathcal{V}_{L,N},
\end{equation}
where $\mathcal{V}_{L,N}$ is an $L\times L$ matrix with entries
\begin{equation}
\left(\mathcal{V}_{L,N}\right)_{ij} =
\begin{cases}
d(u_j) u_j^{2i-1} & i=1,\ldots,N
\\[4pt]
a(u_j) u_j^{2i-3} & i=N+1,\ldots,L.
\end{cases}
\end{equation}
Here, the functions $a(u)=a(u;\xi_1,\ldots,\xi_M)$ and
$d(u)=d(u;\xi_1,\ldots,\xi_M)$ are 
\begin{equation}\label{a-and-d}
a(u)=\frac{\alpha^M}{\Delta^M}\prod_{j=1}^{M}
\left(\frac{u}{\xi_j}-\frac{\xi_j}{u}\right),\qquad
d(u)=\frac{1}{\alpha^M}\prod_{j=1}^{M}\frac{u}{\xi_j}.
\end{equation}
\end{theorem}

The second result concerns the partition function 
in the homogeneous limit.

\begin{theorem}\label{Th2}
The partition function of the homogeneous model can be given 
in terms of an $(L-N)\times (L-N)$ Hankel determinant,
\begin{multline}\label{Zhom1}
Z = (-1)^{\frac{(L-N)(L-N-1)}{2}} 
\prod_{i=1}^{L-N}
\frac{M!(M+i-1)!}{(M-N)!(M+L-N-1)!(N+i-1)!}
\\  \times
\frac{\alpha^{M(L-2N)}}{\Delta^{(L-N)(M-N)}}
x^{\frac{LM}{2}-\frac{(L-N)(L-N-3)}{2}}
\\ \times
\det_{1\leq i,j\leq L-N}
\bigg[(x\pd_x)^{i+j-2}
\frac{(x-1)^{M+L-2N-1}}{x^{M+1}}
\\ \times
\Ftwoone{-N}{L-N-1}{-M}{x}
\bigg],
\end{multline}
or in terms of $N\times N$ determinant,
\begin{multline}\label{Zhom2}
Z = 
\prod_{i=1}^{N}\frac{(L+M-2N)!(M-N)!}{(M-N)!(L-i)!(M-i)!}
\frac{\alpha^{M(L-2N)}}{
\Delta^{(L-N)(M-N)}}
\\ \times
x^{-\frac{L(M-2N)}{2}-\frac{N(N+1)}{2}} 
(x-1)^{ML}
\det_{1\leq i,j\leq N}
\bigg[(x\pd_x)^{i+j-2}
\frac{x^{M-L+1}}{(x-1)^{M+L-2N+1}}
\\ \times
\Ftwoone{-L+N+1}{-L+N}{-L-M+2N}{1-x}
\bigg],
\end{multline}
where $x\equiv u^2$. 
\end{theorem}

We prove Theorem~\ref{Th1} in Sect.~2 and Theorem~\ref{Th2} 
in Sect.~3. 

In Sect.~3, basing on Theorem 2, we also show that
the partition function is a $\tau$-function 
of rational solutions of the sixth Painlev\'e 
equation, where the role of the time variable is 
played by $x$.

\section{Determinant formulas for the inhomogeneous model}

In this section we formulate the five-vertex model in the framework of 
the Quantum Inverse Scattering method (QISM) \cite{KBI-93}. 
The main result here concerns matrix elements of products of $A$- 
or $D$-operators between off-shell Bethe states. The assertion of Theorem \ref{Th1} 
follows from the expressions for these matrix elements. 

\subsection{The partition function as a matrix element}

We begin with giving description of main objects of the QISM \cite{KBI-93} 
related to the five-vertex. Consider a vector space $\mathbb{C}^2$ 
and denote its basis vectors as spin-up and spin-down states 
\begin{equation}\label{updown}
\ket{\uparrow}=
\begin{pmatrix}
1 \\ 0	
\end{pmatrix},
\qquad
\ket{\downarrow}=
\begin{pmatrix}
0 \\ 1	
\end{pmatrix}.
\end{equation}
To each vertical and horizontal line of the square lattice
associate the vector space $\mathbb{C}^2$. We adopt the convention 
that the up and right arrows in the arrow language correspond to 
projection on the spin-up state, and the down and left arrows correspond to
projection on the spin-down state.

\begin{figure} 	
\centering
\input{fig-Lop}
\caption{Definition of the operator $L_{\mu, k}(u_\mu,\xi_k)$ associated to the 
intersection of $\mu$th vertical line (counted from the right) and 
$k$th horizontal line (counted from the top). Here, the arrows indicate directions 
at which the operator acts (the location of ``in'' indices)}
\label{fig-Lop}
\end{figure}
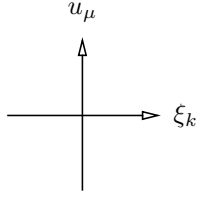

To the vertex lying at the intersection the $\mu$th vertical
and $k$th horizontal lines, associate an operator $L_{\mu, k}=L_{\mu, k}(u_\mu,\xi_k)$, 
which acts non-trivially in the direct product of two spaces:  
``vertical'' space $\mathcal{V}_\mu=\mathbb{C}^2$ (associated with the $\mu$th 
vertical line) and the ``horizontal'' space  $\mathcal{H}_k=\mathbb{C}^2$ 
(associated with the $k$th horizontal line). 
In our construction below, the ``vertical'' spaces play the role of 
``auxiliary'' spaces and the ``horizontal'' spaces play the role of ``quantum'' spaces; 
note that in the QISM one usually makes this 
identification the other way around. 
For a convenience, we label the former spaces by Greek letters (e.g., $\mu, \nu=1,\ldots,L$) 
and the latter ones by Latin letters (e.g., $k,l=1,\ldots,M$). 
The operator $L_{\mu, k}$
describes the Boltzmann weights of the indicated vertex
and its action is shown in Fig.~\ref{fig-Lop} (recall that the vertical lines 
are enumerated from the right to left and 
the horizontal ones from the top to bottom). 

Using the convention between arrow states and the projections to the 
spin-up and spin-down states introduced above, one finds that for the five-vertex model 
\begin{multline}\label{Lop5v}
L_{\mu, k}
=w_1\left(\frac{1+\tau^z_\mu}{2}\right)
\left(\frac{1+\sigma^z_k}{2}\right)
+w_3
\left(\frac{1-\tau^z_\mu}{2}\right)
\left(\frac{1+\sigma^z_k}{2}\right)
\\
+w_4
\left(\frac{1+\tau^z_\mu}{2}\right)
\left(\frac{1-\sigma^z_k}{2}\right)
+w_5\,\tau^-_\mu
\sigma^+_k
+w_6\,\tau^+_\mu
\sigma^-_k,
\end{multline}
where $\tau_\mu^i$ and $\sigma_k^i$ ($i=+,-,z$) denote operators acting as 
Pauli matrices in $\mathcal{V}_\mu$ and $\mathcal{H}_k$, respectively, and 
as identity operators in other spaces.   
More explicitly, in the  tensor product basis $\ket{\uparrow}\otimes\ket{\uparrow}$, 
$\ket{\uparrow}\otimes\ket{\downarrow}$, 
$\ket{\downarrow}\otimes\ket{\uparrow}$, 
$\ket{\downarrow}\otimes\ket{\downarrow}$ in $\mathcal{V}_\mu\otimes \mathcal{H}_k$, 
one has 
\begin{equation}\label{Lexplicit}
L_{\mu, k}=
\begin{pmatrix}
w_1 & 0 & 0 & 0
\\
0 & w_4 & w_6 & 0 
\\
0 & w_5 & w_3 & 0 
\\
0 & 0 & 0 & 0
\end{pmatrix}_{[\mathcal{V}_\mu\otimes \mathcal{H}_k]}.
\end{equation}
Here, $w_i=w_i(u_k/\xi_\mu)$, and the functions $w_i(u)$, $i=1,3,4,5,6$,
are defined in \eqref{weights}. 

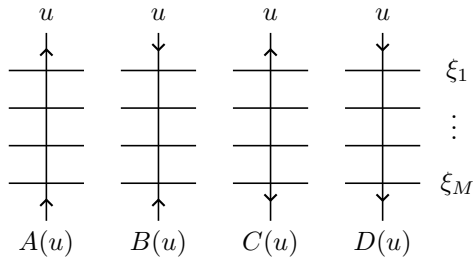
\begin{figure} 
\centering
\input{fig-ABCD}
\caption{Graphical representation of the 
operators---elements of the monodromy matrix $T_\mu(u)=T_\mu(u;\xi_1,\ldots,\xi_M)$}
\label{fig-ABCD}
\end{figure}

As the next step, we introduce the quantum monodromy matrix 
which is defined as an order product of the L-operators along 
the auxiliary direction, which in our construction corresponds to 
a vertical line, 
\begin{equation}\label{eq:T=}
T_\mu (u_\mu;\xi_1,\ldots,\xi_M) 
= L_{\mu, M} (u_\mu, \xi_M)\cdots
L_{\mu, 2} (u_\mu, \xi_2)
L_{\mu, 1} (u_\mu, \xi_1).
\end{equation}
It acts non-trivially in $\mathcal{V}_\mu\otimes \mathcal{H}$ where $\mathcal{H}$ 
is the total quantum space, $\mathcal{H}=\otimes_{k=1}^M\mathcal{H}_k$. In the
spin basis \eqref{updown} in $\mathcal{V}_\mu$ it is has the form of 
an $2\times 2$ matrix with the operator entries acting in $\mathcal{H}$,
\begin{equation}
T_\mu(u)
= \begin{pmatrix}
A (u) & B (u) \\
C (u) & D (u) \\
\end{pmatrix}_{[\mathcal{V}_\mu]},
\end{equation}
where the operators $A(u)=A(u,\xi_1,\ldots,\xi_M)$, etc, are independent 
of the number of the column, $\mu$. 
These operators admit useful graphical representation as columns of the
lattice with the arrows on the top and bottom external vertical edges fixed, 
see Fig.~\ref{fig-ABCD}.

Let us now consider the partition function of the five-vertex model 
shown in Fig.~\ref{fig-LxMlattice}. 
The boundary conditions on the right (respectively, left) boundaries
are described by the ket (bra) state ``all spins up'',
\begin{equation}
\ket{\Omega} = \ket{\ua}^{\otimes M}.
\end{equation}
We will refer to this state as the vacuum state. 

To describe the boundary conditions on the top and bottom boundaries, 
let us turn to the graphical interpretation of the elements of the monodromy matrix 
shown in Fig.~\ref{fig-ABCD}. Let us first consider the case $L=2N$. In this case, 
the arrow states on the top and bottom boundary are all opposite to each other;
specifically, the partition function is given as 
the following vacuum matrix element of the 
product of $C$- and $B$-operators:
\begin{equation}\label{OCBO}
Z = \bra{ \Omega }
\prod_{j=N+1}^{L} C (u_j)
\prod_{j=1}^{N} B (u_j)
\ket{\Omega},
\qquad 
L=2N.
\end{equation}
In other words, it is given as a scalar product of 
two $N$-particle off-shell Bethe states. 
Clearly, the general case differs from \eqref{OCBO} in that there are products of  
$A$- or $D$-operators inserted between off-shell Bethe states.  

To consider the general case, let us denote 
\begin{equation}\label{newnm}
m=|L-2N|,\qquad n=\min (N,L-N).
\end{equation}
Note that \eqref{newnm} simply means that if $L-2N\geq 0$, then $n=N$ and $m=L-2N$, and if 
$L-2N\leq 0$, then $n=L-N$ and $m=2N-L$. 
Here $m$ and $n$ are introduced such that $m$   
gives the number of the $A$- or $D$-operators and $n$ gives
the number of particles in the off-shell Bethe states (the number of 
the $C$- and $B$-operators), $L=2n+m$. 

Now, we can write the partition function for the general case. We have
\begin{equation}\label{CAB}
Z = \bra{ \Omega }
\prod_{j=n+m+1}^{L} 
C (u_j)
\prod_{j=n+1}^{n+m} 
A (u_j)
\prod_{j=1}^{n} B (u_j)
\ket{\Omega},\qquad 
L\geq 2N,
\end{equation}
and, similarly,
\begin{equation}\label{CDB}
Z = \bra{ \Omega }
\prod_{j=n+m+1}^{L} 
C (u_j)
\prod_{j=n+1}^{n+m} 
D (u_j)
\prod_{j=1}^{n} B (u_j)
\ket{\Omega},\qquad 
L \leq 2N.
\end{equation}
For $m=0$, the expressions in \eqref{CAB} and \eqref{CDB} must be  
understood as formula \eqref{OCBO}.

We finish here by mentioning that $\ket{\Omega}$ 
is an eigenvector of the $A$- and $D$-oper\-ators,
\begin{equation}
A(u)\ket{\Omega} 
=a(u)\ket{\Omega},
\qquad
D(u)\ket{\Omega} = d(u)\ket{\Omega}.
\end{equation}
In parametrization \eqref{weights} for the eigenvalues, we have 
\begin{equation}\label{audu}
a(u)=\frac{\alpha^M}{\Delta^M}\prod_{j=1}^{M}
\left(\frac{u}{\xi_j}-\frac{\xi_j}{u}\right),
\qquad
d(u)=\frac{u^M}{\alpha^M}\prod_{j=1}^{M}
\frac{1}{\xi_j}.
\end{equation}
Note that the results of this section, as far as they concern matrix elements
of operators described by  \eqref{CAB} or \eqref{CDB}, are in fact valid for 
arbitrary functions $a(u)$ and $d(u)$. 

\subsection{Yang-Baxter algebra}

The operator $L_{\mu, k}$ satisfies
the intertwining relation  
\begin{equation}\label{RLL=LLR}
R_{\mu,\nu} (u, v) L_{\mu, k}(u,\xi) L_{\nu, k} (v,\xi)
= L_{\mu, k}(v,\xi) L_{\nu, k} (u,\xi)  R_{\mu,\nu} (u, v),
\end{equation}
where $R_{\mu,\nu}(u, v)$ acts non-trivially in $\mathcal{V}_\mu\otimes\mathcal{V}_\nu$
and in the basis of tensor product of two spaces reads
\begin{equation}\label{Rmatrix}
R_{\mu,\nu} (u, v) 
=
\begin{pmatrix}
f(v,u) & 0 & 0 & 0\\
0 & g(v,u) & 1 & 0\\
0 & 0 & g(v,u) & 0\\
0 & 0 & 0 & f(v,u)\\
\end{pmatrix}_{[\mathcal{V}_\mu\otimes\mathcal{V}_\nu]}.
\end{equation}
The entries are given by the functions
\begin{equation}\label{fg-func} 
f(v,u)=\frac{\Delta v^2}{v^2-u^2}, \qquad 
g(v,u)=\frac{\Delta vu}{v^2-u^2},
\end{equation}
where $\Delta$ is the parameter of the weights, see \eqref{weights}.
It is useful to mention, that the ``L-operator'' 
\eqref{Lop5v} (or \eqref{Lexplicit}) is not a unique solution of the relation 
\eqref{RLL=LLR} with the given ``R-matrix'' \eqref{Rmatrix}. It corresponds to
the finite-dimensional quantum space solution 
in the case where $\mathcal{H}_k=\mathbb{C}^2$ \cite{BN-97}. 
Various solutions of the ``RRL relation'' \eqref{RLL=LLR} with 
the R-matrix \eqref{Rmatrix} are given
in Appendix A.

Relation \eqref{RLL=LLR} implies that the monodromy matrix 
satisfies the intertwining relation 
\begin{equation}\label{RTT=TTR}
R_{\mu,\nu} (u,v)
T_\mu (u) T_\nu (v)
= T_\mu (v) T_\nu (u) 
R_{\mu,\nu} (u,v).
\end{equation}
Using \eqref{RTT=TTR} we get the commutation relations for 
the entries of the monodromy matrix $T$. The most important are
\begin{equation}\label{XX=XX}
X(u) X(v) =X(v) X(u),\qquad X\in\{A,B,C,D\},
\end{equation}
and
\begin{align}\label{AB}
A(v)B(u)&=f(v,u)B(u)A(v)+g(u,v)B(v)A(u), 
\\ \label{DB}
D(v)B(u)&=f(u,v)B(u)D(v)+g(v,u)B(v)D(u),
\\ \label{CB}
C(u)B(v)&=g(u,v)\left[A(u)D(v)-A(v)D(u)\right],
\end{align}
where the functions $f(v,u)$ and $g(v,u)$ are defined in \eqref{fg-func}.

In what follows we 
need standard formulas which describe 
the action of the $A$- and $D$-operators on an
off-shell Bethe state.  
\begin{lemma}\label{BBOmega}
The following relations are valid 
\begin{equation}\label{ABBOmega}
A(u_{n+1})\prod_{j=1}^{n} B(u_j)\ket{\Omega}
=\sum_{i=1}^{n+1} a(u_i)
\frac{g(u_i,u_{n+1})}{f(u_i,u_{n+1})}
\prod_{\substack{j=1\\ j\ne i}}^{n+1}
f(u_i,u_j)\prod_{\substack{j=1\\ j\ne i}}^{n+1}	B(u_j)
\ket{\Omega}
\end{equation}
and 
\begin{equation}\label{DBBOmega}
D(u_{n+1})\prod_{j=1}^{n} B(u_j)\ket{\Omega}
= \sum_{i=1}^{n+1} d(u_i)
\frac{g(u_{n+1},u_i)}{f(u_{n+1},u_i)}
\prod_{\substack{j=1\\ j\ne i}}^{n+1} f(u_j,u_i)
\prod_{\substack{j=1\\ j\ne i}}^{n+1} B(u_j)\ket{\Omega},
\end{equation}
where $a(u)$ and $d(u)$ are vacuum eigenvalues of the 
operators $A(u)$ and $D(u)$, respectively.
\end{lemma}
\begin{proof}
Consider, for example, the case of the $A$-operator. 
Due to the standard technique of the algebraic Bethe Anzatz,  
acting with the $A$-operator on the off-shell state by relation
\eqref{AB}, one gets
\begin{multline}
A(w)\prod_{j=1}^{n} B(u_j)\ket{\Omega}
=a(w)\prod_{j=1}^{n} f(w,u_j) \prod_{j=1}^{n} B(u_j)\ket{\Omega}
\\ 
+\sum_{i=1}^{n} 
a(u_i) g(u_i, w) 
\prod_{\substack{j=1\\j \ne i}}^{n} f(u_i,u_j) B(w)
\prod_{\substack{j=1\\j \ne i}}^{n} B(u_j)
\ket{\Omega}.
\end{multline}
Setting $w=u_{n+1}$ and taking into account that $f(u,v)/g(u,v)=u/v$ and 
hence $f(u,u)/g(u,u)=1$, we can combine 
the terms in the right hand side into a single sum, thus obtaining 
\eqref{ABBOmega}. 

The proof of \eqref{DBBOmega} based 
on relation \eqref{DB} is similar.
\end{proof}

To illustrate a peculiarity of the Yang-Baxter algebra governed by the R-matrix 
\eqref{Rmatrix}, it is useful to mention symmetry properties of the products 
of operators standing in \eqref{CAB} and \eqref{CDB}. In a rather general setup, 
we have the following. 

\begin{proposition}
The operators 
\begin{equation}\label{CAB-lmn}
\prod_{j=1}^{m}v_{j}^{-1}
\prod_{j=1}^{l} C (u_j)
\prod_{j=1}^{m} 
A (v_j)
\prod_{j=1}^{n} B (w_j)
\end{equation}
and 
\begin{equation}\label{CDB-lmn}
\prod_{j=1}^{m}v_{j}
\prod_{j=1}^{l} 
C (u_j)
\prod_{j=1}^{m} 
D (v_j)
\prod_{j=1}^{n} B (w_j)
\end{equation}
are totally symmetric in permutations of all the $l+m+n$ 
parameters in the set $\{u\}\cup\{v\}\cup\{w\}$.
\end{proposition}
\begin{proof}
Due to commutation relations \eqref{XX=XX}, the operators \eqref{CAB-lmn} and \eqref{CDB-lmn}
are obviously symmetric with respect to permutations of the elements within each set 
$\{u\}$, $\{v\}$, and $\{w\}$. To study merging of these sets, let us 
first consider the special case $m=0$, which corresponds to the operator 
$\prod_{j=1}^{l} C (u_j)\prod_{j=1}^{n} B (w_j)$.
The relation $C(u)B(w)=C(w)B(u)$ which is relation \eqref{CB1}, see the full list 
of the commutation relations in Appendix A, implies that since there are symmetries in 
permutations of $u$'s and $w$'s, one can mix elements of the sets 
$\{u\}$ and $\{w\}$ in an arbitrary way. Thus, the operator 
$\prod_{j=1}^{l} C (u_j)\prod_{j=1}^{n} B (w_j)$ is totally symmetric with respect permutations
of the elements of the set $\{u\}\cup\{w\}$. 

Consider now the operator \eqref{CAB-lmn}. Denote $\tilde A(v)\equiv v^{-1} A(v)$, and consider 
operators  $\prod_{j=1}^{l} C (u_j) \prod_{j=1}^{m} \tilde A (v_j)$ and 
$\prod_{j=1}^{m} \tilde A (v_j) \prod_{j=1}^{n} B (w_j)$. Again, each of these two operators 
is totally symmetric with respect to permutations of all its variables. This is due, 
respectively, to relation \eqref{CA1}, which reads $C(u)\tilde A(v)= C(v)\tilde A(u)$, 
and relation \eqref{AB1}, which reads $\tilde A(v) B(w)= \tilde A(w) B(v)$. Hence, 
the operator $\prod_{j=1}^{l}C (u_j)\prod_{j=1}^{m}\tilde A (v_j)\prod_{j=1}^{n} B (w_j)$,
which is exactly \eqref{CAB-lmn}, is totally symmetric in all its variables. 

The proof in the case of the operator \eqref{CDB-lmn} is essentially similar. 
Here, one can denote $\tilde D(v) \equiv v D(v)$ and use relations 
\eqref{CD1} and \eqref{DB1}, which reads $C(u)\tilde D(v)= C(v)\tilde D(u)$ and 
$\tilde D(v) B(w)= \tilde D(w) B(v)$, respectively.
\end{proof}

\subsection{Matrix elements of products of $A$- and $D$-operators}

In this section we state and prove the main result about 
matrix elements of the products of $A$- or $D$-operators. 
We introduce the notation
\begin{equation}
\begin{split}
S_{L,n}^{[A]}=\bra{\Omega}
\prod_{j=1}^{n} C(u_{n+m+j})
\prod_{j=1}^{m} A(u_{n+j})
\prod_{j=1}^{n} B(u_j)
\ket{\Omega},
\\ 
S_{L,n}^{[D]}=\bra{\Omega}
\prod_{j=1}^{n} C(u_{n+m+j})
\prod_{j=1}^{m} D(u_{n+j})
\prod_{j=1}^{n} B(u_j)
\ket{\Omega},
\end{split}
\end{equation}
where the superscript reminds that $S_{L,n}^{[A]}$ and 
$S_{L,n}^{[D]}$ are matrix elements of the product 
of the $A$- and $D$-operators, respectively, and $L= 2n+m$. 
These matrix elements can be given in terms of determinants of 
$L\times L$ matrices.

\begin{theorem}\label{SAD}
Let $A(u)$, $B(u)$, $C(u)$, $D(u)$ be operators satisfying commutation relations 
\eqref{AB}--\eqref{CB}, and $a(u)$ and $d(u)$ be 
vacuum eigenvalues of the operators $A(u)$ and $D(u)$, respectively,
\begin{equation}
A(u)\ket{\Omega} = a(u)\ket{\Omega},
\qquad
D(u)\ket{\Omega} = d(u)\ket{\Omega},
\end{equation}
then the following representations are valid:
\begin{equation}\label{ProdA}
S_{L,n}^{[A]}
=\Delta^{(n+m)n}
\prod_{j=1}^{m} u_{n+j}\prod_{1\leq i<j\leq L}\frac{1}{u_j^2-u_i^2}
\det\mathcal{V}_{L,n}
\end{equation}
and
\begin{equation}\label{ProdD}
S_{L,n}^{[D]}
= \Delta^{(n+m)n}
\prod_{j=1}^{m} u_{n+j}^{-1}\prod_{1\leq i<j\leq L}\frac{1}{u_j^2-u_i^2}
\det \mathcal{V}_{L,n+m},
\end{equation}
where $\mathcal{V}_{L,n}$ is an $L\times L$ matrix, $L=2n+m$, with the entries
\begin{equation}\label{VLn}
\left(\mathcal{V}_{L,n}\right)_{ij} =
\begin{cases}
d(u_j) u_j^{2i-1} & i=1,\ldots,n
\\[4pt]
a(u_j) u_j^{2i-3} & i=n+1,\ldots,L.
\end{cases}
\end{equation}
\end{theorem}

The remaining part of this subsection is devoted to the proof. 
We consider in detail the case of matrix elements of the product of
$A$-operators; the proof in the case of matrix elements of the product of
$D$-operators is essentially similar. The proof is by induction in $n$ and
$m$, separately, and thus splits naturally into two induction steps.

We first verify the base case, that corresponds to $n=1$ and $m=0$, that is, 
to the scalar product of $1$-particle off-shell states, 
$S_{2,1}^{[A]}=S_{2,1}^{[D]}=:S_{1}$. 
In this case, using commutation relation \eqref{CB}, we have
\begin{align}
S_{1}
&
=\bra{\Omega}C(u_2)B(u_1)\ket{\Omega}
\\ &
= g(u_2,u_1)\bra{\Omega} \left[A(u_2)D(u_1)-A(u_1)D(u_2)\right]\ket{\Omega}.
\end{align}
Recalling that $\ket{\Omega}$ is eigenvector of both $A$- and $D$-operators, we get
\begin{align}
S_{1}&=g(u_2,u_1) \left[a(u_2)d(u_1)-a(u_1)d(u_2)\right]
\\ & 
=\frac{\Delta}{u_2^2-u_1^2} 
\begin{vmatrix}
d(u_1) u_1 & d(u_2) u_2 \\
a(u_1) u_1 & a(u_2) u_2
\end{vmatrix}.
\end{align}
This is exactly the right hand side of \eqref{ProdA} 
and \eqref{ProdD} for $n = 1$ and $m = 0$.

Let us turn now to considering the induction steps. 
In the first induction step we prove that \eqref{ProdA}
holds for 
some $n$ and $m$, assuming that it is valid for
$n$ and $m-1$. Writing 
\begin{equation}
S_{L,n}^{[A]}=\bra{\Omega}
\prod_{j=1}^{n} C(u_{n+m+j})
\prod_{j=2}^{m} A(u_{n+j})\cdot A(u_{n+1})
\prod_{j=1}^{n} B(u_j)
\ket{\Omega},
\end{equation}
and appying Lemma~\ref{BBOmega} which gives the result of action with the
$A$-operator on the off-shell Bethe state, see \eqref{ABBOmega}, we obtain the relation
\begin{equation}\label{SLn-A}
S_{L,n}^{[A]}=\sum_{i=1}^{n+1} 
a(u_i)\frac{g(u_i,u_{n+1})}{f(u_i,u_{n+1})}
\prod_{\substack{j=1 \\ j\ne i}}^{n+1} f(u_i,u_j)
S_{L-1,n}^{[A]}(\setminus u_i).
\end{equation}
Here, $S_{L-1,n}^{[A]}(\setminus u_i)$ denotes the matrix element 
with one $A$-operator less and the variable $u_i$ excluded from the 
set of all variables entering the operators. Our aim now is to show that 
the expression standing in \eqref{ProdA} solves recurrence relation 
\eqref{SLn-A}. 

Namely, substituiting 
\eqref{ProdA} for $S_{L-1,n}^{[A]}(\setminus u_i)$ into the right-hand side of 
\eqref{SLn-A}, we get 
\begin{multline}
S_{L,n}^{[A]}
=\Delta^{(n+m-1)n}
\sum_{i=1}^{n+1} 
a(u_i)\frac{g(u_i,u_{n+1})}{f(u_i,u_{n+1})}
\prod_{\substack{j=1\\j\ne i}}^{n+1}
f(u_i,u_j)
\\ \times
\prod_{j=2}^{m} u_{n+j}
\prod_{\substack{1\leq j<l\leq L\\ j,l\ne i}}\frac{1}{u_l^2-u_j^2}
\det \mathcal{V}_{L-1,n} (\setminus u_i),
\end{multline}
where $\mathcal{V}_{L-1,n} (\setminus u_i)$ denotes the $(L-1)\times (L-1)$ matrix obtained from 
$\mathcal{V}_{L,n}$ by removing the last row and the $i$-th column.
Using the explicit form of the functions $f(u,v)$ and $g(u,v)$, see \eqref{fg-func}, 
we obtain
\begin{equation}
S_{L,n}^{[A]}
= \Delta^{(n+m)n}
\sum_{i=1}^{n+1} 
\frac{a(u_i)}{u_i}
\prod_{\substack{j=1\\ j\ne i}}^{n+1}\frac{u_i^2}{u_i^2-u_j^2}
\prod_{j=1}^{m} u_{n+j}
\prod_{\substack{1\leq j<l\leq L\\ j,l\ne i}}\frac{1}{u_l^2-u_j^2}
\det \mathcal{V}_{L-1,n} (\setminus u_i).
\end{equation}
Simplifying further, we get
\begin{multline}
S_{L,n}^{[A]}
=\Delta^{(n+m)n}
\prod_{j=1}^{m} u_{n+j}
\prod_{1\leq i<j\leq L}\frac{1}{u_j^2-u_i^2}
\\ \times
\sum_{i=1}^{n+1} (-1)^{i-1} a(u_i) u_i^{2n-1}
\prod_{j=n+2}^{L} (u_i^2-u_j^2)
\det \mathcal{V}_{L-1,n} (\setminus u_i).
\end{multline}
Let us now look closer at the sum above. 
This is nothing but a minor expansion of the determinant of the 
matrix 
\begin{equation}
\big(\widetilde{\mathcal{V}}_{L,n}\big)_{ij} =
\begin{cases}
d(u_j)u_j^{2i-1} &i=1,\ldots,n 
\\[4pt]
a(u_j)u_j^{2i-3} &i=n+1,\ldots,L-1 
\\
a(u_j)u_j^{2n-1}\displaystyle\prod_{k=n+2}^{L}(u_j^2-u_k^2) & i=L.
\end{cases}
\end{equation}
Note that the last row of this matrix is equal to $a(u_j)u_j^{2n-1}$ times 
a polynomial in $u_j^2$ with the 
coefficients independent on the column number $j$. 
Therefore all terms in the last row can be removed by
subtraction of the successive rows, except the term 
$a(u_j)u_j^{2L-3}$. Hence, 
\begin{equation}
\det \widetilde{\mathcal{V}}_{L,n} =
\det\mathcal{V}_{L,n},
\end{equation}
that completes the proof of the
first induction step.

Let us now consider the second induction step. 
Namely, we prove that \eqref{ProdA} holds for 
some $n$ and $m=0$, assuming that it is valid for
$n-1$ and $m=1$.

Consider the scalar product of the off-shell Bethe states, 
$S_{2n,n}^{[A]}=S_{2n,n}^{[D]}=:S_{n}$.
Using commutation relation \eqref{CB}, we first write
\begin{align}
S_{n}
&=\bra{\Omega}\!\!\!
\prod_{j=n+2}^{2n}\!\!\! C(u_j)\cdot
C(u_{n+1})B(u_{n})
\cdot\prod_{j=1}^{n-1} B(u_j)\ket{\Omega}
\\ 
&=g(u_{n+1},u_{n})\Bigg\{
\bra{\Omega}\!\!\!
\prod_{j=n+2}^{2n}\!\!\! C(u_{j})
A(u_{n+1})D(u_{n})
\prod_{j=1}^{n-1} B(u_j)\ket{\Omega}
- (u_{n} \leftrightarrow u_{n+1})\Bigg\}.
\end{align}
Next, using again Lemma~\ref{BBOmega}, now in that part which gives  
the result of action with the $D$-operator on the off-shell Bethe state, 
see \eqref{DBBOmega}, 
we obtain the relation
\begin{multline}\label{S2nn}
S_{n}	
=g(u_{n+1},u_{n})\Bigg\{
\sum_{i=1}^{n}d(u_i)\frac{g(u_n,u_i)}{f(u_n,u_i)}
\prod_{\substack{j=1\\ j\ne i}}^{n}
f(u_j,u_i)S_{2n-1,n-1}^{[A]}(\setminus u_i)
\\ 
-(u_{n} \leftrightarrow u_{n+1})\Bigg\}.
\end{multline}
Here, $S_{2n-1,n-1}^{[A]}(\setminus u_i)$ denotes the matrix element 
of the operator $A(u_{n+1})$, in which in the set of its  parameters the parameter
$u_{i}$ is absent, or, more precisely, it depends on the parameters
$u_{2n},\ldots,u_{n+2},u_{n+1},u_{n},\ldots,u_{i+1},u_{i-1},\ldots,u_{1}$, in that order.

Now, applying the induction step, that is, substituting 
the expression for $S_{2n-1,n-1}^{[A]}$ in \eqref{S2nn}, 
and using expression for the $f$- and $g$-functions,  
we literally get 
\begin{multline}
S_{n}=
\frac{\Delta u_{n+1} u_{n}}{u_{n+1}^2-u_{n}^2}
\Bigg\{
\sum_{i=1}^{n}
d(u_i)
\frac{u_i}{u_n}
\prod_{\substack{j=1\\ j\ne i}}^{n} 
\frac{\Delta u_j^2}{u_j^2-u_i^2}
\\\times
\Delta^{(n-1)n}u_{n+1}
\prod_{\substack{1\leq l<k\leq 2n\\ l,k\ne i}}
\frac{1}{u_k^2-u_l^2}
\det\mathcal{V}_{2n-1,n-1}(\setminus u_i)
-\left(u_{n}\leftrightarrow u_{n+1}\right)\Bigg\}.
\end{multline}
Simplifying and extracting from the sum a total antisymmetric factor, we obtain
\begin{multline}\label{lastS2nn}
S_{n}=
\frac{\Delta^{n^2}}{u_{n+1}^2-u_{n}^2}
\prod_{j=1}^{n+1} u_j^2
\prod_{1\leq l<k\leq 2n}\frac{1}{u_k^2-u_l^2}
\\\times
\Bigg\{\sum_{i=1}^{n}(-1)^{i-1}
\frac{d(u_i)}{u_i}
\prod_{j=n+1}^{2n}
(u_j^2-u_i^2)
\det\mathcal{V}_{2n-1,n-1}(\setminus u_i)
+\left(u_{n}\leftrightarrow u_{n+1}\right)\Bigg\}.
\end{multline}
Consider the expression in the braces in \eqref{lastS2nn}. Using the antisymmetry 
of the factor $\det\mathcal{V}_{2n-1,n-1}(\setminus u_i)$ with respect to exchange 
of $u_n$ and $u_{n+1}$ for $i\ne n, n+1$, we can rewrite this expression in the form
\begin{align}\label{longsums}
&{}\sum_{i=1}^{n-1}(-1)^{i-1}
\frac{d(u_i)}{u_i}
(u_{n+1}^2-u_i^2)\prod_{j=n+2}^{2n}
(u_j^2-u_i^2)
\det\mathcal{V}_{2n-1,n-1}(\setminus u_i)
\\ &\quad
+(-1)^{n-1}\frac{d(u_n)}{u_n}(u_{n+1}^2-u_n^2)
\prod_{j=n+2}^{2n}
(u_j^2-u_n^2) 
\det\mathcal{V}_{2n-1,n-1}(\setminus u_n)
\\ &\quad
+\left(u_{n}\leftrightarrow u_{n+1}\right)
\\ &
= (u_{n+1}^2-u_n^2) \sum_{i=1}^{n+1}
(-1)^{i-1}\frac{d(u_i)}{u_i}
\prod_{j=n+2}^{2n}
(u_j^2-u_i^2)
\det\mathcal{V}_{2n-1,n-1}(\setminus u_i)
\\ &
= (u_{n+1}^2-u_n^2) \det \mathcal W_{2n,n},
\end{align}
where $W_{2n,n}$ denotes a $2n\times 2n$ matrix with the entries
\begin{equation}\label{Wmatrix}
\left(\mathcal{W}_{2n,n}\right)_{ij} =
\begin{cases}
\displaystyle\frac{d(u_j)}{u_j}
\prod_{l=n+2}^{2n} (u_l^2-u_j^2) & i = 1 
\\
d(u_j)u_j^{2i-3} & i=2,\ldots,n
\\
a(u_j)u_j^{2i-5} & i=n+1,\ldots,2n. 
\end{cases}
\end{equation}

Clearly, in the determinant of the 
matrix $\mathcal{W}_{2n,n}$ we can change the entries of the first row ($i=1$) 
by subtracting, with suitable 
coefficients, the entries of the next $n-1$ successive rows ($i=2,\ldots,n$)
such that in the first row at the $j$-th column we leave with the term $u_j^{-1}d(u_j)$ times 
the factor $\prod_{l=n+2}^{2n} u_l^2$. Moving this factor out of the determinant
and giving the $j$-th column the factor $u_j^2$ allows us 
to obtain under the determinant exactly 
the matrix $\mathcal{V}_{2n,n}$, or, in other words, 
we arrive at the identity
\begin{equation}\label{prodW=V}
\prod_{j=1}^{n+1} u_j^{2} \det \mathcal{W}_{2n,n} =\det\mathcal{V}_{2n,n}.
\end{equation}

Finally, plugging \eqref{longsums} into \eqref{lastS2nn}
with  \eqref{prodW=V} taken into account, exactly gives the expected result 
of the second induction step, namely, the expression for $S_{n}$
given by \eqref{ProdA} or \eqref{ProdD} at $m=0$ and $L=2n$. This 
concludes the proof of Theorem \ref{SAD}.

\section{Determinant formulas for the homogeneous model}

In this section we consider the partition function in the 
homogeneous limit. We obtain various representations 
in terms of Hankel determinants, in particular, that appearing in 
the assertion of Theorem \ref{Th2}. We also 
show that the partition function is the $\tau$-function of the 
sixth Painlev\'e equation, in a special case where it corresponds to 
classical solutions, related to the Gauss hypergeometric function.

\subsection{The homogeneous limit}

Let us study the result for the partition function, 
expressed by formula \eqref{ZdetVLN}, in the case where 
the Boltzmann weights of the model are site-independent, 
that corresponds to taking the limits 
$u_1,\ldots,u_L\to u$ and $\xi_1,\ldots,\xi_M\to 1$. 
Note that only the first limit (in $u$'s) is singular in 
\eqref{ZdetVLN}, while the second one (in $\xi$'s) 
is trivial since these variable enter only the functions 
$a(u)$ and $d(u)$. In fact, we can treat 
the first limit even without specifying explicit form of these functions.

To obtain an expression for this limit, we apply the relation
\begin{equation}\label{homlimVLN}
\lim_{u_1,\ldots,u_L\to u}
\prod_{1\leq l<k\leq L}\frac{1}{u_k^2-u_l^2}
\det\mathcal{V}_{L,N}
=\det\mathcal{Q}_{L,N}
\end{equation}
where the matrix $\mathcal{Q}_{L,N}$ has entries
\begin{equation}
\left(\mathcal{Q}_{L,N}\right)_{ij}=
\begin{cases}
\displaystyle\frac{1}{(j-1)!}
\left(\frac{\partial}{\partial u^2}\right)^{j-1}
d(u)u^{2i-1}
& i=1,\ldots,N
\\
\displaystyle
\frac{1}{(j-1)!}
\left(\frac{\partial}{\partial u^2}\right)^{j-1}
a(u)u^{2i-3}
&i=N+1,\ldots,L.
\end{cases}
\end{equation}
This can be proved by expanding the elements of $\mathcal{Q}_{L,N}$ in Taylor series 
and subtracting recursively columns of the determinant. 
Note that the same result is applicable to 
the matrix elements \eqref{ProdA} and \eqref{ProdD}. 

As for the partition function, we thus find 
that \eqref{ZdetVLN} in the homogeneous limit becomes
\begin{equation}\label{Zhomfirst}
Z=\Delta^{N(L-N)} u^{L-2N} \det\mathcal{Q}_{L,N}.
\end{equation}
It turns out that it can also be given in terms of Hankel 
determinants, either of size $N\times N$ or $(N-L)\times (N-L)$.
The following generalizes the result  of \cite{P-16} for the $L=2N$ case. 

\begin{proposition} 
For the partition function $Z$ 
in the homogeneous limit the following formulas are valid: 
\begin{multline}\label{Z=detL-N}
Z=(-1)^{\frac{(L-N)(L-N-1)}{2}}
\Delta^{N(L-N)}
[d(u)]^{L}
u^{2(L-N)}
\\ \times
\det_{1\leq i,j\leq L-N}
\left[\left.
\frac{1}{s!}
\left(\pdv{}{u^2}\right)^{s}
\frac{a(u)}{d(u)}u^{2N-2}
\right|_{s=i+j-1-L+2N}
\right]
\end{multline}
and 
\begin{multline}\label{Z=detN}
Z=(-1)^{NL+\frac{N(N+1)}{2}}
\Delta^{N(L-N)}
[a(u)]^{L}
u^{2N(L-1)}
\\ \times
\det_{1\leq i,j \leq N}
\left[
\left.
\frac{1}{s!}
\left(\pdv{}{u^2}\right)^{s}
\frac{d(u)}{a(u)}u^{2-2N}
\right|_{s=i+j-1+L-2N}
\right].
\end{multline}
\end{proposition}

\begin{proof}
Introduce notation
\begin{equation}\label{tilde-a-d}
\tilde{d}=\tilde{d}(u^2)\equiv d(u)u,\qquad
\tilde{a}=\tilde{a}(u^2)\equiv a(u)u^{2N-1}.
\end{equation}
We first note that, by subtraction of rows, it can be shown that
\begin{equation}\label{detq=detwtQ}
\det \mathcal{Q}_{L,N}=\det \widetilde{\mathcal{Q}}_{L,N},
\end{equation}
where 
\begin{equation}\label{wtQmatrix}
\big(\widetilde{\mathcal{Q}}_{L,N}\big)_{ij}=
\begin{cases}
\displaystyle\frac{\tilde{d}^{(j-i)}}{(j-i)!} 
& i=1,\ldots,N
\\[10pt]
\displaystyle\frac{\tilde{a}^{(N+j-i)}}{(N+j-i)!}
&i=N+1,\ldots,L.
\end{cases}
\end{equation}
The matrix \eqref{wtQmatrix} can be regarded as having a block form
\begin{equation}\label{wtQblock}
\widetilde{\mathcal{Q}}_{L,N} =  
\begin{pmatrix}
\mathcal{A} & \mathcal{B} \\
\mathcal{C} & \mathcal{D} \\
\end{pmatrix},
\end{equation}
where either $\mathcal{A}$ or $\mathcal{C}$ are upper 
triangular square matrices. 

To prove \eqref{Z=detL-N} we take the first option, where 
$\mathcal{A}$ and $\mathcal{D}$ 
are matrices of the sizes $N\times N$ and 
$(L-N)\times (L-N)$, respectively. 
For the determinant of the block matrix we use
\begin{equation}
\det
\begin{pmatrix}
\mathcal{A} & \mathcal{B} \\
\mathcal{C} & \mathcal{D} \\
\end{pmatrix}
=\det\mathcal{A}\cdot
\det\left(\mathcal{D}-\mathcal{C}\mathcal{A}^{-1}\mathcal{B}
\right).
\end{equation}
Since the matrix $\mathcal{A}$ is upper triangular, 
and, furthermore, of Teoplitz form, it can be easily inverted, 
\begin{equation}
\left(\mathcal{A}^{-1}\right)_{jk} 
=\frac{1}{(k-j)!} 
\left(\frac{1}{\tilde{d}} \right)^{(k-j)}.
\end{equation}
Indeed, this can be verified directly
\begin{multline}
\sum_{k=1}^{N}
\left(\mathcal{A}^{-1}\right)_{jk}\mathcal{A}_{kl}
=\sum_{k=1}^{N}\frac{1}{(k-j)!}\left(\frac{1}{\tilde{d}}\right)^{(k-j)}
\frac{\tilde{d}^{(l-k)}}{(l-k)!}
\\
=\frac{1}{(s-j)!}\sum_{k=j}^{l}\binom{l-j}{k-j}
\left(\frac{1}{\tilde{d}}\right)^{(k-j)}
\tilde{d}^{(l-k)}
=\frac{1}{(l-j)!}\left(\pdv{}{u^2}\right)^{l-j}1
=\delta_{jl}.
\end{multline}
In a similar way, we obtain
\begin{equation}
\sum_{l=1}^{N}
\mathcal{C}_{il}\left(\mathcal{A}^{-1}\right)_{lk}
=\sum_{l=1}^{N}
\frac{\tilde{a}^{(l-i)}}{(l-i)!}\frac{1}{(k-l)!} 
\left(\frac{1}{\tilde{d}} \right)^{(k-l)}
=\frac{1}{(k-i)!}
\left(\frac{\tilde{a}}{\tilde{d}}\right)^{(k-i)},
\end{equation}
and therefore
\begin{multline}\label{CA-1B}
\left(\mathcal{C}\mathcal{A}^{-1}\mathcal{B}\right)_{ij}
=\sum_{k=1}^{N}
\frac{1}{(k-i)!}
\left(\frac{\tilde{a}}{\tilde{d}}\right)^{(k-i)}\frac{\tilde{d}^{(N+j-k)}}{(N+j-k)!}
\\
=\frac{1}{(N+j-i)!}	
\sum_{k=i}^{N}\binom{N+j-i}{k-i}\left(\frac{\tilde{a}}{\tilde{d}}\right)^{(k-i)}
\tilde{d}^{(N+j-k)}.
\end{multline}
Writing the entries of the matrix $\mathcal{D}$ as a sum,
\begin{equation}
\mathcal{D}_{ij}=\frac{\tilde{a}^{(N+j-i)}}{(N+j-i)!}=
\frac{1}{(N+j-i)!}	
\sum_{k=0}^{N+j-i}\binom{N+j-i}{k}\left(\frac{\tilde{a}}{\tilde{d}}\right)^{(k)}
\tilde{d}^{(N+j-i-k)},
\end{equation}
we see that \eqref{CA-1B} exactly gives the terms which correspond to 
$k=0,\ldots,N-i$ elements of this sum, and thus
\begin{multline}\label{D-CAB}
\left(\mathcal{D}-\mathcal{C}\mathcal{A}^{-1}\mathcal{B}
\right)_{ij}
=\frac{1}{(N+j-i)!}
\sum_{k=N-i+1}^{N+j-i}
\binom{N+j-i}{k}
\left(\frac{\tilde{a}}{\tilde{d}}\right)^{(k)}\tilde{d}^{(N+j-i-k)}
\\
=\sum_{l=0}^{j-1}
\frac{1}{(N+j-i-l)!l!}
\left(\frac{\tilde{a}}{\tilde{d}}\right)^{(N+j-i-l)}\tilde{d}^{(l)}.
\end{multline}
The last expression in \eqref{D-CAB} shows, that in evaluating of the determinant 
one can recursively remove all terms in the sum, except the $l=0$ term, 
by the column subtraction,  
\begin{equation}
\det\left(\mathcal{D}-\mathcal{C}\mathcal{A}^{-1}\mathcal{B}\right)
=\det_{1\leq i,j \leq L-N}
\left[\frac{\tilde{d}}{(N+j-i)!}
\left(\frac{\tilde{a}}{\tilde{d}}\right)^{(N+j-i)}\right].
\end{equation}
Finally, we reflect the matrix here by changing $i\to L-N+1-i$ and 
move the factor $\tilde{d}$ out of the determinant, thus obtaining 
\begin{equation}
\det\widetilde{\mathcal{Q}}_{L,N}=(-1)^{\frac{(L-N)(L-N-1)}{2}}
\tilde{d}^{L} 
\det_{1\leq i,j\leq L-N}
\left[\frac{1}{s!}\left(\frac{\tilde{a}}{\tilde{d}}\right)^{(s)}
\bigg|_{s=i+j-1-L+2N}\right].
\end{equation}
Recalling notation \eqref{tilde-a-d} and relation \eqref{detq=detwtQ}, from 
\eqref{Zhomfirst} we arrive at formula \eqref{Z=detL-N}.

Let us now consider representation \eqref{Z=detN}. To derive
it, we take the second option in \eqref{wtQblock}, where 
$\mathcal{B}$ and $\mathcal{C}$ are matrices of the 
sizes $N\times N$ and $(L-N)\times (L-N)$, respectively, and
use the relation
\begin{equation}
\det
\begin{pmatrix}
\mathcal{A} & \mathcal{B} \\
\mathcal{C} & \mathcal{D} \\
\end{pmatrix}
=(-1)^{\dim \mathcal{B} \dim\mathcal{C}}
\det\mathcal{C}\det\left(\mathcal{B}-\mathcal{A}\mathcal{C}^{-1}\mathcal{D}\right).
\end{equation}
From this relation, using the fact that $\mathcal{C}$ is an upper triangular 
Toeplitz matrix, in the essentially similar manner, we obtain
\begin{equation}
\det\widetilde{\mathcal{Q}}_{L,N}=
(-1)^{NL+\frac{N(N+1)}{2}}\tilde{a}^{L}
\det_{1\leq i,j \leq N}
\left[\frac{1}{s!}
\bigg(\frac{\tilde{d}}{\tilde{a}}
\bigg)^{(s)}
\bigg|_{s=i+j-1+L-2N}\right].
\end{equation}
Again, using \eqref{Zhomfirst}, \eqref{tilde-a-d} and \eqref{detq=detwtQ}, 
we thus arrive at \eqref{Z=detN}.
\end{proof}

\subsection{Another Hankel determinant formulas}

Representations \eqref{Z=detL-N} and \eqref{Z=detN} express 
the partition function in terms of Hankel determinants, which are related, 
as shown below, to random matrix integrals over closed contours in the complex plane. 
It turns out that these integrals can be effectively evaluated, leading to 
another Hankel determinants which are implicitly related to 
certain discrete random matrix models. It is this result that is stated in 
Theorem \ref{Th2}, and we prove it here, extending calculations of 
\cite{P-16} for the case $L=2N$ to the general situation.  

To derive these formulas, we need to 
resort to explicit expressions \eqref{audu} 
for the functions $a(u)$ and $d(u)$ for the homogeneous model ($\xi_j=1$).
Representation \eqref{Z=detL-N} now reads
\begin{multline}\label{Zhomdet1}
Z = (-1)^{\frac{(L-N)(L-N-1)}{2}} 
\frac{\alpha^{M(L-2N)}}{
\Delta^{(L-N)(M-N)}}
x^{L-N+\frac{ML}{2}}
\\ \times
\det_{1\leq i,j\leq L-N}
\left[\frac{1}{s!}
\pd_x^{s}
\frac{(x-1)^M}{x^{M-N+1}}
\bigg|_{s=i+j-1-L+2N}
\right]
\end{multline}
and \eqref{Z=detN} reads
\begin{multline}\label{Zhomdet2}
Z = (-1)^{LN+\frac{N(N+1)}{2}} 
\frac{\alpha^{M(L-2N)}}{
\Delta^{(L-N)(M-N)}}
x^{N(L-1)-\frac{ML}{2}} (x-1)^{ML}
\\ \times
\det_{1\leq i,j\leq N}
\left[\frac{1}{s!}
\pd_x^{s}
\frac{x^{M-N+1}}{(x-1)^M}
\bigg|_{s=i+j-1+L-2N}
\right],
\end{multline}
where $x\equiv u^2$ and $\pd_x\equiv \partial/\partial x$. The derivation involves several steps,
and we consider both representations \eqref{Zhomdet1} and \eqref{Zhomdet2}
in parallel.

On the first step we obtain multiple integral representations.
We first use the Cauchy formula to rewrite entries as contour integrals,
\begin{equation}\label{Cauchy}
\frac{1}{s!}
\pd_x^{s}
f(x)
= \oint_{C_x}\frac{f(z)}{(z-x)^{s+1}}
\frac{\dd z}{2\pi\ii},
\end{equation}
where $C_x$ denotes a simple positive-oriented contour around 
the point $z=x$, and next apply the following identity. 
\begin{lemma}
Let $x$ and $y$ be some parameters, and $\gamma$ is some contour in the complex plane, 
then 
\begin{multline}\label{det=ooint}
\det_{1\leq i,j\leq n}
\left[\int_\gamma (z-x)^{n-i}(z-y)^{n-j}
f(z)
\frac{\dd z}{2\pi\ii}
\right]
\\ = \frac{1}{n!}
\int_\gamma\dots\int_\gamma
\prod_{1\le i<j\le n}
\left(z_i-z_j\right)^2
\prod_{j=1}^n
f(z_j)
\frac{\dd z_j}{2\pi\ii},
\end{multline}
where $f(z)$ is a function such that all integrals exist, 
and which may also depend on $x$ and $y$, 
$f(z)=f(z;x,y)$.
\end{lemma}

Indeed, choosing $x=y$ and $\gamma=C_x$, we can rewrite 
the determinants in \eqref{Zhomdet1} and \eqref{Zhomdet2} 
as multiple contour integrals, where, in the case 
of \eqref{Zhomdet1}, we have $n=L-N$ and 
\begin{equation}\label{fzone}
f(z)=\frac{(z-1)^M}{z^{M-N+1}(z-x)^{L}},
\end{equation}
and, in the case of \eqref{Zhomdet2}, we have $n=N$ and
\begin{equation}\label{fztwo}
f(z)=\frac{z^{M-N+1}}{(z-x)^{L}(z-1)^{M}}.
\end{equation}
As a result, we get representations in terms of random matrix model integrals 
on a closed contour in the complex plane. 

On the second step we turn back to a determinant form 
using \eqref{det=ooint} in the reverse direction, where we introduce the parameter
$y$ by a replacement $(z-1)^M\mapsto (z-y)^M$
in functions \eqref{fzone} and \eqref{fztwo}. It is assumed that
$y$ takes values outside of the interior of the contour $C_x$, and 
after calculations one has to take $y=1$. As a result, for the determinant in 
\eqref{Zhomdet1}, we get
\begin{multline}
\det_{1\leq i,j\leq L-N}
\left[\frac{1}{s!}
\pd_x^s
\frac{(x-1)^M}{x^{M-N+1}}
\bigg|_{s=i+j-1-L+2N}\right]
\\ 
= \det_{1\leq i,j\leq L-N}
\left[
\oint_{C_x}	
\frac{(z-y)^{M+L-N-j}}{z^{M-N+1}(z-x)^{N+i}}
\frac{\dd z}{2\pi\ii}\right]
\bigg|_{y=1},
\end{multline}
and, for the determinant in \eqref{Zhomdet2}, we get
\begin{multline}
\det_{1\leq i,j\leq N}
\left[\frac{1}{s!}
\pd_x^s
\frac{x^{M-N+1}}{(x-1)^M}
\bigg|_{s=i+j-1+L-2N}
\right]
\\ 
= \det_{1\leq i,j\leq N}
\left[
\oint_{C_x}	
\frac{z^{M-N+1}}{(z-x)^{L-N+i}(z-y)^{M-N+j}}
\frac{\dd z}{2\pi\ii}
\right]
\bigg|_{y=1}.
\end{multline}
In these expressions, 
one can treat 
the terms $(z-x)^{-i}$ and $(z-y)^{-j}$ 
in the integrals as derivatives with respect to $x$ and $y$, respectively. 
This gives
\begin{multline}\label{NewDet1}
\det_{1\leq i,j\leq L-N}
\left[\frac{1}{s!}
\pd_x^s
\frac{(x-1)^M}{x^{M-N+1}}
\bigg|_{s=i+j-1-L+2N}\right]
\\
=(-1)^{\frac{(L-N)(L-N-1)}{2}}\prod_{i=1}^{L-N}
\frac{N!(M+i-1)!}{(N+i-1)!(M+L-N-1)!}
\\  \times
\det_{1\leq i,j\leq L-N}
\left[
\pd_x^{i-1}\pd_y^{j-1}
\oint_{C_x}\frac{(z-y)^{M+L-N-1}}{z^{M-N+1}(z-x)^{N+1}}
\frac{\dd z}{2\pi\ii}
\right]\bigg|_{y=1}
\end{multline}
and 
\begin{multline}\label{NewDet2}
\det_{1\leq i,j\leq N}
\left[\frac{1}{s!}
\pd_x^s
\frac{x^{M-N+1}}{(x-1)^M}
\bigg|_{s=i+j-1+L-2N}
\right]
=\prod_{i=1}^{N}\frac{(L-N)!(M-N)!}{(L-i)!(M-i)!}
\\ \times
\det_{1\leq i,j\leq N}
\left[
\pd_x^{i-1}\pd_y^{j-1}
\oint_{C_x}	
\frac{z^{M-N+1}}{(z-x)^{L-N+1}(z-y)^{M-N+1}}
\frac{\dd z}{2\pi\ii}
\right]\bigg|_{y=1}.
\end{multline}

On the third step, we rewrite these determinants 
as Hankel determinants, using the following identity.

\begin{lemma}\label{Lm-dets}
For a homogeneous function $h(x,y)$, 
the following identity is valid:
\begin{equation}\label{deteqdet}
\det_{1\leq i,j\leq n}
\left[\pd_x^{i-1}\pd_y^{j-1}
h(x,y)\right]\big|_{y=1}
=\left(-\frac{1}{x}\right)^{\frac{n (n-1)}{2}}
\det_{1\leq i,j\leq n}
\left[(x\pd_x)^{i+j-2}h(x,1)\right].
\end{equation}
\end{lemma}
\begin{proof}
We first note that, by subtracting rows and columns, one can 
easily see the validity of the following two identities:
\begin{equation}
\det_{1\leq i,j\leq n} 
\left[\pd_x^{i-1}\pd_y^{j-1}
h(x,y)\right]
=\det_{1\leq i,j\leq n}
\left[\frac{1}{x^{i-1}y^{j-1}}
(x\pd_x)^{i-1}
(y\pd_y)^{j-1}
h(x,y)\right],
\end{equation}
and
\begin{equation}
\det_{1\leq i,j\leq n} 
\left[(x\pd_x)^{i-1}
(x\pd_y)^{j-1}
h(x,y)
\right]
=\det_{1\leq i,j\leq n}
\left[x^a y^b
(x\pd_x)^{i-1}
(y\pd_y)^{j-1}
\frac{h(x,y)}{x^a y^b}
\right].
\end{equation}
Since $h(x,y)$ is a homogeneous function of a degree, say, 
$\nu$, that is $h(x,y) = y^{\nu}h(x/y,1)$, and taking into account 
that $y\partial_y h(x/y,1)=-x\partial_x h(x/y,1)$, we have 
\begin{align}
\det_{1\leq i,j\leq n} 
\left[\pd_x^{i-1}\pd_y^{j-1}
h(x,y)\right]
&=\frac{1}{(xy)^{n(n-1)/2}}
\det_{1\leq i,j\leq n}
\left[(x\pd_x)^{i-1}(y\pd_y)^{j-1}
h(x,y)\right],
\\  &
=\frac{y^{\nu n}}{(xy)^{n(n-1)/2}}
\det_{1\leq i,j\leq n}
\left[(x\pd_x)^{i-1}
(y\pd_y)^{j-1}
h(x/y,1)\right],
\\  &
=\frac{(-1)^{n(n-1)/2}y^{\nu n}}{(xy)^{n(n-1)/2}}
\det_{1\leq i,j\leq n}
\left[(x\pd_x)^{i+j-2}
h(x/y,1)\right].
\end{align}
Setting now $y=1$, we arrive at \eqref{deteqdet}.
\end{proof}

Furthermore, the function $h(x)\equiv h(x,1)$ can be evaluated by 
Cauchy formula \eqref{Cauchy} in each case; for \eqref{NewDet1}, 
we get
\begin{align}
h(x) 
&
=\oint_{C_x}\frac{(z-1)^{M+L-N-1}}{z^{M-N+1}(z-x)^{N+1}}
\frac{\dd z}{2\pi\ii}
\\ &
=\frac{1}{N!} 
\pd_x^N
\frac{(x-1)^{M+L-N-1}}
{x^{M-N+1}},
\end{align}
and for \eqref{NewDet2}, we get
\begin{align}
h(x)
&
=\oint_{C_x}	
\frac{z^{M-N+1}}{(z-x)^{L-N+1}(z-1)^{M-N+1}}
\frac{\dd z}{2\pi\ii}
\\ &
=\frac{1}{(L-N)!} 
\pd_x^{L-N}
\frac{x^{M-N+1}}
{(x-1)^{M-N+1}}.
\end{align}
Summarizing, we thus obtain that 
\eqref{Zhomdet1} can be rewritten as follows
\begin{multline}\label{Zhom1new}
Z=(-1)^{\frac{(L-N)(L-N-1)}{2}}
\prod_{i=1}^{L-N}
\frac{(M+i-1)!}{(N+i-1)!(M+L-N-1)!}
\frac{\alpha^{M(L-2N)}}{\Delta^{(L-N)(M-N)}}
\\ \times
x^{\frac{ML}{2}-\frac{(L-N)(L-N-3)}{2}}
\det_{1\leq i,j\leq L-N}
\left[
(x\pd_x)^{i+j-2}
\pd_x^N
\frac{(x-1)^{M+L-N-1}}
{x^{M-N+1}}
\right],
\end{multline}
and \eqref{Zhomdet2} as follows
\begin{multline}\label{Zhom2new}
Z = (-1)^{\frac{(L-N)N}{2}} 
\prod_{i=1}^{N}\frac{(M-N)!}{(L-i)!(M-i)!}
\frac{\alpha^{M(L-2N)}}{
\Delta^{(L-N)(M-N)}}
\\ \times
\frac{(x-1)^{ML}}{x^{\frac{L(M-2N)}{2}+\frac{N(N+1)}{2}} }
\det_{1\leq i,j\leq N}
\left[
(x\pd_x)^{i+j-2}
\pd_x^{L-N}
\frac{x^{M-N+1}}
{(x-1)^{M-N+1}}
\right].
\end{multline}

At the fourth, final, step, we express the functions 
in the Hankel determinants standing under the symbol 
$(x\pd_x)^{i+j-2}$ in terms of the Gauss hypergeometric function. 
One can use the formula (see, e.g., \cite{B-53}, \S~2.8, (17))
\begin{equation}\label{derivate}
\pd_x^n\frac{1}{(x-1)^a x^b}
=\frac{(b)_n}{(x-1)^{a+n} x^{b+n}}
\Ftwoone{-n}{-n-a-b+1}{-n-b+1}{x},
\end{equation}
where $(b)_n$ is a Pochhammer symbol and $b\ne 0,-1,\ldots,-n+1$. 
In the case of the function in \eqref{Zhom1new}, we have
\begin{multline}\label{h-func1}
\pd_x^N
\frac{(x-1)^{M+L-N-1}}
{x^{M-N+1}}
=\frac{M!}{(M-N)!}
\frac{(x-1)^{M+L-2N-1}}{x^{M+1}}
\\  \times
\Ftwoone{-N}{L-N-1}{-M}{x}. 
\end{multline}
In the case of the function in \eqref{Zhom2new}, we get
\begin{multline}\label{h-func2}
\pd_x^{L-N}
\frac{x^{M-N+1}}
{(x-1)^{M-N+1}}
=(-1)^{L-N}\frac{(M+L-2N)!}{(M-N)!}
\frac{x^{M-L+1}}{(x-1)^{M+L-2N+1}}
\\ \times
\Ftwoone{-L+N+1}{-L+N}{-L-M+2N}{1-x}.
\end{multline}
Here, we have used the analogue of \eqref{derivate} obtained 
by the change $x\mapsto 1-x$ since the limitation in \eqref{derivate} 
become sensible if $L>M+1$; it can be overcome 
if the argument of the Gauss hypergeometric function is chosen to be, say, $1-x$ 
instead of $x$. 

Plugging \eqref{h-func1} in \eqref{Zhom1new} and \eqref{h-func2} in \eqref{Zhom2new}
one gets \eqref{Zhom1} and \eqref{Zhom2}, respectively.
This proves Theorem~\ref{Th2}.

\subsection{Connection with the sixth Painlev\'e equation}

Explicit expressions for the functions in the Hankel determinants 
make it possible to investigate more deeply the dependence of the 
partition function as a function of $x$. It turns out that 
the partition function appears to be  
a $\tau$-function of the sixth Painlev\'e equation. 

To show this, we consider first representation \eqref{Zhom1new} and
apply Euler transformation of Gauss hypergeometric function,
\begin{equation}
\Ftwoone{a}{b}{c}{x}=(1-x)^{-a}\Ftwoone{a}{c-b}{c}{\frac{x}{x-1}},
\end{equation}
to \eqref{h-func1}, that yields
\begin{equation}
\Ftwoone{-N}{L-N-1}{-M}{x}=
(-1)^N(x-1)^N \Ftwoone{-N}{-M-L+N+1}{-M}{\frac{x}{x-1}}. 
\end{equation}
Ignoring an overall factor in $Z$ independent of $x$, 
we then get 
\begin{multline}
Z\sim 
x^{\frac{ML}{2}-\frac{(L-N)(L-N-3)}{2}}
\det_{1\leq i,j\leq L-N}
\bigg[
(x\pd_x)^{i+j-2}
\frac{(x-1)^{M+L-N-1}}{x^{M+1}}
\\ \times
\Ftwoone{-N}{-M-L+N+1}{-M}{\frac{x}{x-1}}
\bigg].
\end{multline}
Let us make change of the variable 
\begin{equation}
x=\frac{t}{t-1}
\end{equation}
and rewrite the partition function in the form  
\begin{multline}\label{Zsimt1}
Z\sim
\left[(t-1)^Lt^{L-2N}\right]^{-\frac{M}{2}}
\left[t(t-1)\right]^{-\frac{1}{2}(b_3+b_4)}t^{-\gamma n/2}
(t-1)^{\gamma n/2}
\\ \times	
\det_{1\leq i,j \leq n}
\left[\delta_t^{i+j-2}
t^A(t-1)^B
\Ftwoone{b_1+b_4}{1-b_1+b_4}{1+b_2+b_4}{t}
\right],
\end{multline}
where $\delta_t\equiv t(t-1)\pd_t$, and where
we have introduced the notation 
\begin{equation}
n=L-N,\quad
A=-M-1,\quad
B=-L+N+2,\quad
\gamma=A-B,
\end{equation}
and 
\begin{equation}\label{b-params}
\begin{split}
b_1&=\frac{1}{2}(L+M)-N,\\
b_2&=\frac{1}{2}(L-M)-1,\\
b_3&=\frac{1}{2}(L-M),\\
b_4&=-\frac{1}{2}(L+M).
\end{split}
\end{equation}
Comparison of formula \eqref{Zsimt1} with the expression for the $\tau$-function 
of the rational solutions of the sixth Painleve equation \cite{FW-04} shows that 
\begin{equation}
\tau(t)\sim \left[(t-1)^Lt^{L-2N}\right]^{-\frac{M}{2}} Z.
\end{equation}
According to the general theory of the sixth Painleve equation \cite{JM-81,O-87}, 
the function 
\begin{equation}
\sigma(t)=t(t-1) \pdv{\log \tau(t)}{t} +(b_1 b_3+b_1b_4+b_3b_4)
-\frac{1}{2}\sum_{1\leq j<k \leq 4}b_j b_k 
\end{equation}
satisfies the equation
\begin{equation}
\sigma'\big(t(t-1)\sigma''\big)^2
+\big(\sigma'[2\sigma+(1-2t)\sigma']+b_1 b_2 b_3 b_4\big)^2 
=\prod_{j=1}^{4}\big(\sigma'+b_j^2\big)
\end{equation}
with the parameters $b_1,\ldots,b_4$ defined in \eqref{b-params}. 

Essentially similarly, one can write expression \eqref{Zhom2new} in the form 
\begin{multline}
Z\sim
\left[(t-1)^Lt^{L-2N}\right]^{-\frac{M}{2}}
\left[t(t-1)\right]^{-\frac{1}{2}(\tilde b_3+\tilde b_4)}t^{-\tilde\gamma \tilde n/2}
(t-1)^{\tilde \gamma \tilde n/2}
\\ \times	
\det_{1\leq i,j \leq \tilde n}
\left[\delta_t^{i+j-2}
t^{\tilde A}(t-1)^{\tilde B}
\Ftwoone{\tilde b_1+\tilde b_4}{1-\tilde b_1+\tilde b_4}{1+\tilde b_2+\tilde b_4}{t}
\right],
\end{multline}
where 
\begin{equation}
\tilde n=N,\qquad
\tilde A=M-L+1,\qquad
\tilde B=L-N,\qquad
\tilde\gamma =\tilde A- \tilde B,
\end{equation}
and 
\begin{equation}\label{b-params2}
\begin{split}
\tilde b_1&=-\frac{1}{2}(L+M)+N,\\ 
\tilde b_2&=-\frac{1}{2}(L-M)+1,\\
\tilde b_3&=\frac{1}{2}(L+M),\\
\tilde b_4&=-\frac{1}{2}(L-M).
\end{split}
\end{equation}
Comparison \eqref{b-params} with \eqref{b-params2} shows that 
$\tilde b_1=-b_1$, $\tilde b_2=-b_1$, $\tilde b_3=-b_4$, and $\tilde b_4= -b_3$, 
and hence the two sets of parameters \eqref{b-params} and \eqref{b-params2} 
define essentially the same equation for the $\sigma$-function.  

In order to obtain the equation in the variable $x$, one has to make the change 
$t=x/(x-1)$. This is a symmetry transformation \cite{O-87,FW-04}, 
which maps the set of parameters $\{b_1,b_2,b_3,b_4\}$ into another 
set of parameters, say, $\{\nu_1,\nu_2,\nu_3,\nu_4\}$. 
More precisely, we have the following.

\begin{proposition}
The $\sigma$-function 
\begin{equation}
\sigma(x)=x(x-1) \pdv{\log Z}{x}-A x+ B
\end{equation} 
where $Z=Z(x)$ is a partition function given by \eqref{Zhom1} or \eqref{Zhom2}, 
and
\begin{equation}
A=\frac{LM}{2}+\frac{(N-1)^2}{4},\qquad
B=\frac{(N+1)(L+M-2N)}{4}+\frac{N^2-M}{2},
\end{equation}
satisfies the  sixth Painlev'e equation in its $\sigma$-form
\begin{equation}
\sigma'\big(x(x-1)\sigma''\big)^2
+\big(\sigma'[2\sigma+(1-2x)\sigma']+\nu_1 \nu_2 \nu_3 \nu_4\big)^2 
=\prod_{j=1}^{4}\big(\sigma'+\nu_j^2\big),
\end{equation}
with the set of parameters
\begin{equation}
\{\nu_1,\nu_2,\nu_3,\nu_4\}=
\left\{M-\frac{N-1}{2},-L+\frac{N+1}{2},\frac{N+1}{2},\frac{N-1}{2}\right\}.
\end{equation}
\end{proposition}
A detailed discussion of the rational solutions of the sixth Painleve equation
can be found, e.g., in \cite{M-01}.

\section*{Acknowledgments}

The authors are indebted to N. M. Bogoliubov, F. Colomo, and N. Reshetikhin
for stimulating discussions. This work is supported in part 
by the Russian Science Foundation, grant \#18-11-00297. 

\appendix
\section{Solutions of the RLL relation}

Here, we consider some solutions of the RLL relation \eqref{RLL=LLR} with the R-matrix
given by \eqref{Rmatrix} and \eqref{fg-func}. To simplify notations, we 
consider here the L-operator as a one-site monodromy matrix, 
\begin{equation}
L(u)=
\begin{pmatrix}
A(u) & B(u)  \\ C(u) & D(u)
\end{pmatrix},
\end{equation}
and write the RLL relation in the form
\begin{equation}
R (u, v) \left(L(u) \otimes L(v)\right)
= \left(L(v) \otimes L(u)\right)  R (u, v),
\end{equation}
where $R(u,v)\equiv R_{\mu,\nu}(u,v)$ is given by \eqref{Rmatrix} and \eqref{fg-func}.  
We use the following shorthand notation: $A\equiv A(u)$, etc, and 
$f\equiv f(v,u)$ and $g\equiv g(v,u)$. The prime denotes the change $u\leftrightarrow v$, 
e.g., $A'\equiv A(v)$ and $g'\equiv g(u,v)$.

The 16 commutation relations contained in the RLL-relation then read:
\begin{subequations}
\begin{align}\label{AA}
\left[A, A'\right] &=0,
\\ \label{AB1}
A B' & =\frac{u}{v}A' B,
\\  \label{AB2}
A'B&=fBA'+g'B'A,
\\  \label{BB}
\left[B,B'\right]&=0,
\\ \label{CA2}
CA'&=fA'C+g'AC',
\\ \label{CB=AD}
CB'&=g\left(A'D-AD'\right),
\\ \label{DA2}
\left[D,A'\right]&=g\left(B'C-BC'\right),
\\ \label{DB2}
DB'&=fB'D+g'BD',
\\ \label{CA1}
CA'&=\frac{v}{u}C'A,
\\ \label{CB1}
CB'&=C'B,
\\ \label{CB=DA}
C'B&=g\left(DA'-D'A\right),
\\ \label{DB1}
DB'&=\frac{v}{u}D'B,
\\ \label{CC}
\left[C,C'\right]&=0,
\\ \label{CD1}
CD'&=\frac{u}{v}C'D,
\\ \label{CD2}
C'D&=fDC'+g'D'C,
\\ \label{DD}
\left[D,D'\right]&=0.
\end{align}
\end{subequations}
Here, we have used the fact that $f/g=v/u$.

\subsection{An infinite dimensional ``bosonic'' solution}

We first consider a solution in which $A$, $B$, $C$, and $D$ are 
operators acting in the Fock space of a single boson (quantum oscillator), spanned 
by the Fock number states $\ket{n}$, $n\in \mathbb{Z}_{\geq0}$, satisfying $\braket{n}=1$. 
In the basis $\ket{0}, \ket{1},\dots,$ these operators are
semi-infinite dimensional matrices.

We assume that: (\emph{i}) $B$ and $C$ are independent of $u$, (\emph{ii})
they are creation and annihilation operators, respectively,
\begin{equation}
B=\sum_{n\geq 0} b_n \ket{n+1}\bra{n}, \qquad
C=\sum_{n\geq 0} c_n \ket{n}\bra{n+1},
\end{equation}
and (\emph{iii}) the operators $A$ and $D$ are diagonal. With these assumptions, 
six relations, namely, \eqref{AA}, \eqref{BB}, \eqref{DA2}, \eqref{CB1}, \eqref{CC}, 
\eqref{DD}, are satisfied automatically. 

To proceed, we restrict ourselves to the situation of general setup where all 
coefficients $b_n$ and $c_n$, $n\in\mathbb{Z}_{\geq 0}$, are nonzero. 
Let us introduce notation for the number and the vacuum projection operators:
\begin{equation}
\hatn =\sum_{n\geq 0} n \ket{n}\bra{n},\qquad
\hat\pi = \ket{0}\bra{0}.
\end{equation}
Considering relations \eqref{AB1}, \eqref{AB2}, \eqref{CA2}, and \eqref{CA1}, 
we find that the operator $A$ has the form
\begin{equation}
A=\alpha_1 u\Delta^{\hatn} +\frac{\alpha_2}{u}\hat \pi
\end{equation}
and, similarly, from \eqref{DB2}, \eqref{DB1}, \eqref{CD1}, and \eqref{CD2}, we obtain
\begin{equation}
D=\frac{\delta_1}{u} \Delta^{\hatn} +\delta_2 u\hat \pi,
\end{equation}
where $\alpha_1$, $\alpha_2$, $\delta_1$, and $\delta_2$ are some parameters. 
From the remaining two relations \eqref{CB=DA} and \eqref{CB=AD} it follows that
\begin{equation}
c_0b_0=(\alpha_1\delta_1-\alpha_2\delta_2)\Delta,
\qquad
c_jb_j=\alpha_1\delta_1 \Delta^{2j+1},\qquad j=1,2,\dots.
\end{equation}

Choosing the overall normalization such that 
\begin{equation}
\alpha_1\delta_1=\Delta^{-1},
\end{equation}
we thus find, that modulo a diagonal similarity transformation, 
the operators $B$ and $C$ are given by 
\begin{equation}
c_0=1\pm\sqrt{\alpha_2\delta_2\Delta},\qquad 
b_0=1\mp\sqrt{\alpha_2\delta_2\Delta},\qquad
c_j=b_j=\Delta^j,\qquad j=1,2,\ldots,
\end{equation}
where $\alpha_2\delta_2\ne 1/\Delta$, to ensure that neither $b_0=0$ nor $c_0=0$.
Removing the remaining one-parameter freedom in rescaling of the rapidity variable, we 
fix completely our solution by setting
\begin{equation}
\alpha_1=\frac{1}{\Delta},\qquad 
\delta_1=1,\qquad 
\alpha_2\equiv-\frac{\alpha}{\Delta},\qquad
\delta_2\equiv \delta, \qquad \alpha\delta\ne -1,
\end{equation}
where $\alpha$ and $\delta$ are new independent parameters. 

Summarizing, we thus have obtained the following expression for the L-operator:
\begin{equation}\label{Lboson}
L(u)=
\begin{pmatrix}
\displaystyle 
u\Delta^{\hatn-1}-\frac{\alpha}{\Delta u}\hat \pi
& 
\hat\varphi^\dagger \big(1\pm\ii\sqrt{\alpha\delta}\hat\pi\big)
\\[6pt]
\big(1\mp\ii\sqrt{\alpha\delta}\hat\pi\big)\hat\varphi
&
\displaystyle 
\frac{1}{u}\Delta^{\hatn}+\delta u\hat \pi
\end{pmatrix},\qquad \alpha\delta\ne -1,
\end{equation}
where $\hat\varphi$ and $\hat\varphi^\dagger$ are bosonic operators:
\begin{equation}
\hat\varphi^\dagger=\sum_{n\geq 0} \Delta^n \ket{n+1}\bra{n}, \qquad
\hat\varphi=\sum_{n\geq 0} \Delta^n \ket{n}\bra{n+1}.
\end{equation}
These operators have the following properties
\begin{equation}
\hat\varphi\hat\varphi^\dagger=\Delta^{2\hatn}, \qquad
\hat\varphi^\dagger\hat\varphi=\Delta^{2\hatn-2}-\Delta^{-2}\hat\pi,
\end{equation}
which imply the commutation relation
\begin{equation}
\hat\varphi\hat\varphi^\dagger-\Delta^2 \hat\varphi^\dagger\hat\varphi=\hat\pi.
\end{equation}
At $\Delta=1$ these operators turn into the so-called 
exponential phase operators studied in solid state physics and quantum optics 
(see, e.g., \cite{BN-97,BIKPT-01} and references therein).

The solution \eqref{Lboson} at $\Delta=1$ and $\delta=0$ reproduces 
(up to the change $u\mapsto u^{-1}$) the L-operator first provided in \cite{BN-97}.
The related non-hermitian bosonic model has applications in 
growth problems as well as it describes enumeration of 
boxed plane partitions \cite{B-05,BM-15,BM-18}.

\subsection{Finite-dimensional solutions}

As noticed in \cite{BN-97}, for the value $\Delta=1$, 
there exists an L-operator expressible in terms of 
operators $S^{+}$ and $S^{-}$ satisfying the relations
\begin{equation}\label{SPM}
S^\pm S^\mp S^\pm=S^\pm,
\qquad 
\left(S^\pm\right)^2=0.  
\end{equation}
For example, these operators can be represented in terms of $(m+1)\times (m+1)$ 
matrices where $S^{-}$ has nonzero entries only on the first column or 
the last row, excluding the diagonal entry, and $S^{+}$ is its transpose,
\begin{equation}\label{findim}
S^{-}
=\begin{pmatrix}
0 & 0 \\ \vec n & 0
\end{pmatrix},
\qquad 
S^{+}=(S^{-})^\mathsf{T},
\qquad
\vec n^2=1,
\end{equation}
where $\vec n$ is an $m$-component column or row vector. 
Here, we consider a generalization of that solution
for arbitrary values of $\Delta$. 
We show
that in fact there exist two solutions, which 
also depend on additional arbitrary parameters, and valid for arbitrary  
realization of algebra \eqref{SPM}, i.e., not necessarily in the form 
of matrices \eqref{findim}. The special case of matrices \eqref{findim} for 
$m=1$, that corresponds to the five-vertex model,
in discussed in detail in the next subsection. 

We start constructing the L-operator with setting $B=S^{-}$ and $C=S^{+}$, that 
fulfill three relations \eqref{BB}, \eqref{CB1}, and \eqref{CC}. Note that with this 
choice we also fix the overall normalization. 

Next, considering relations \eqref{AB1}, \eqref{AB2}, \eqref{CA2}, 
and \eqref{CA1}, we find that 
\begin{equation}\label{Agen}
A=\alpha_1 u Y_1 + \frac{\alpha_2}{u} X_1, 	
\end{equation}
and, similarly, from \eqref{DB2}, \eqref{DB1}, \eqref{CD1}, and \eqref{CD2}, we find that
\begin{equation}\label{Dgen}
D=\frac{\delta_1}{u} Y_2 +\delta_2 u X_2,
\end{equation}
where $\alpha_1$, $\alpha_2$, $\delta_1$, and $\delta_2$ are parameters, and
$X_1$ and $X_2$ are operators satisfying the relations
\begin{equation}
XS^{-}=0, \qquad S^{+}X=0,
\end{equation}
and $Y_1$ and $Y_2$ --- the relations
\begin{equation}
YS^{-}= \Delta S^{-}Y,\qquad  S^{+}Y=\Delta YS^{+}.
\end{equation}
Searching these operators in the form of linear combinations of the operators 
$S^{+}S^{-}$, $S^{-}S^{+}$ and the identity operator $I$, yields for $X$ and $Y$,
modulo an overall normalization, one-parameter solutions
\begin{align}
X&=\frac{1}{1+\beta}S^{+}S^{-}+\frac{\beta}{1+\beta}(I-S^{-}S^{+}),
\\
Y&=\frac{1}{1+\gamma}I +\left(\frac{1}{\Delta}-\frac{1}{1+\gamma}\right)S^{+}S^{-}
+\frac{\gamma}{1+\gamma}S^{-}S^{+},
\end{align}
where $\beta$ and $\gamma$ are parameters, and the operators are normalized 
such that $S^{-}X=YS^{-}=S^{-}$. 

Setting $X_1=X(\beta_1)$, $X_2=X(\beta_2)$, $Y_1=Y(\gamma_1)$, $Y_2=Y(\gamma_2)$, 
where $\beta_1$, $\beta_2$, $\gamma_1$, $\gamma_2$ are some parameters to be determined,
from \eqref{CB=DA} and \eqref{CB=AD} we obtain the following equations 
\begin{equation}\label{3eqs}
\alpha_1\delta_1=0,\qquad \alpha_2\delta_2=-\Delta^{-1}, \qquad \beta_1\beta_2=0.
\end{equation}
From the first equation in \eqref{3eqs} it follows that, structurally, there are two solutions, 
characterized whether $\alpha_1$ or $\delta_1$ vanishes, see \eqref{Agen} and \eqref{Dgen}; 
the third relation splits further these solutions on ``sub-solutions''. 

Consider first the case $\delta_1=0$. Expressing $\alpha_2=-(\delta_2\Delta)^{-1}$, 
and setting $\gamma_1\equiv\gamma$, $Y_1\equiv Y=Y(\gamma)$ we find that 
\begin{equation}
A=\alpha_1 u\, Y -\frac{1}{\Delta\delta_2 u} X_1,\qquad
D=\delta_2 u X_2,\qquad \beta_1\beta_2=0.
\end{equation}
Fixing the scale of $u$ by setting $\delta_2^{-1}=\alpha_1\equiv \alpha$, we thus find 
that the first solution reads
\begin{equation}\label{FirstL}
L(u)=
\begin{pmatrix}
\displaystyle
\alpha u Y - \frac{\alpha}{\Delta u} X_1
& S^{-}
\\[6pt]
S^{+}
& \displaystyle \frac{u}{\alpha} X_2
\end{pmatrix},\qquad \beta_1\beta_2=0.
\end{equation}
In the case of $\alpha_1=0$, we can write, essentially similarly, 
$\delta_2=-(\alpha_2\Delta)^{-1}$, and setting $\gamma_2\equiv \gamma$, 
$Y_2\equiv Y=Y(\gamma)$, we get 
\begin{equation}
A=\frac{\alpha_2}{u} X_1,\qquad
D=\frac{\delta_1}{u}Y-\frac{u}{\alpha_2\Delta} X_2.
\end{equation}
We fix the solution by setting $\alpha_2=\delta_1^{-1}\equiv \alpha$. Hence,
the second solution reads
\begin{equation}\label{SecondL}
L(u)=
\begin{pmatrix}
\displaystyle
\frac{\alpha}{u} X_1
& S^{-}
\\[6pt]
S^{+}
& \displaystyle
\frac{1}{\alpha u}Y- \frac{u}{\alpha \Delta}X_2
\end{pmatrix},\qquad
\beta_1\beta_2=0.
\end{equation}
Note, that we have in total 
three parameters which parametrize the solutions \eqref{FirstL} and 
\eqref{SecondL}: $\alpha$, $\beta$, and $\gamma$, where 
$\beta=\beta_2$, if $\beta_1=0$, or $\beta=\beta_1$, if $\beta_2=0$.

The solution \eqref{SecondL} at $\Delta=1$, 
$\alpha=1$, and $\beta_1=\beta_2=\gamma=0$ (note that $X(0)=S^{+}S^{-}$ and $Y(0)|_{\Delta=1}=I$) 
reproduces (up to the change $u\mapsto u^{-1}$) the L-operator  
considered in \cite{BN-97} (see Eq.~(39) therein).  

As a last comment here, we mention that 
there is a special case of both \eqref{FirstL} and \eqref{SecondL}, 
which corresponds to $\alpha_1=\delta_1=0$ in \eqref{3eqs} and reads 
\begin{equation}\label{ThirdL}
L(u)=
\begin{pmatrix}
\displaystyle
-\frac{1}{\Delta u} X_1
& S^{-}
\\[10pt]
S^{+}
& \displaystyle
u X_2
\end{pmatrix},\qquad
\beta_1\beta_2=0.
\end{equation}
The $L$-operator \eqref{ThirdL} can be obtained from \eqref{FirstL} 
by setting $u\mapsto \alpha u$ and 
taking the limit $\alpha\to 0$, or from \eqref{SecondL}, 
by setting $u\mapsto -\alpha \Delta u$, 
and taking the limit $\alpha\to \infty$.

\subsection{Five- and four-vertex models}

The five-vertex model is described by the finite-dimensional solutions 
in the case of $m=1$. Indeed, in this case, 
\begin{equation}
S^{-}=
\sigma^{-},\qquad
S^{+}=
\sigma^{+},\qquad
X=
\begin{pmatrix}
1 & 0 \\0 & 0
\end{pmatrix},\qquad
Y=
\begin{pmatrix}
\Delta^{-1} & 0 \\0 & 1
\end{pmatrix}.
\end{equation}
The first solution, L-operator \eqref{FirstL}, reads 
\begin{equation}\label{L5V-1}
L(u)=
\begin{pmatrix}
\displaystyle
\alpha\big(u-{u}^{-1}\big)/\Delta & 0 & 0 & 0
\\
0 & \alpha u & 1 & 0
\\
0 & 1 & u/\alpha & 0
\\
0 & 0 & 0 & 0
\end{pmatrix}.
\end{equation}
Comparing this expression with the L-operator of six-vertex model 
as defined as a matrix of Boltzmann weights (see Figs.~\ref{fig-SixVertices} and 
\ref{fig-Lop}, and definition of the $L$-operator before Eqn.~\eqref{Lexplicit})
\begin{equation}\label{L6v}
L=
\begin{pmatrix}
w_1 & 0 & 0 & 0
\\
0 & w_4 & w_6 & 0 
\\
0 & w_5 & w_3 & 0 
\\
0 & 0 & 0 & w_2
\end{pmatrix},
\end{equation}
one obtains that \eqref{L5V-1} corresponds to the 
five-vertex model with weights \eqref{weights}. Note that, from
expression of the weight $w_1$ one can find an interpretation 
of the parameter $\Delta$ in terms of the weights in this model 
as follows:
\begin{equation}\label{Delta-w2}
\Delta=\frac{\alpha}{w_1}\left(u-\frac{1}{u}\right)=
\frac{w_4}{w_1}-\frac{1}{w_1w_3}=\frac{w_3w_4-w_5w_6}{w_1w_3},
\end{equation}
where the weights $w_5$ and $w_6$ are restored due to the overall normalization.

Consider now the second solution, \eqref{SecondL}. We have
\begin{equation}\label{L5V-2}
L(u)=
\begin{pmatrix}
\displaystyle
\alpha/ u & 0 & 0 & 0
\\
0 & 0 & 1 & 0
\\
0 & 1 & (\alpha \Delta)^{-1}\big({u}^{-1}-u\big) & 0
\\
0 & 0 & 0 & (\alpha u)^{-1}
\end{pmatrix}.
\end{equation}
Comparing \eqref{L5V-2} with \eqref{L6v}, we conclude that this solution corresponds to 
the five-vertex model in which $w_4=0$, and 
\begin{equation}\label{Delta-w4}
\Delta= \frac{1}{w_3 \alpha}\left(\frac{1}{u}-u\right)=\frac{w_2}{w_3}-\frac{1}{w_1w_3}=
\frac{w_1w_2-w_5w_6}{w_1 w_3}.
\end{equation}

The five-vertex model in which $w_1=0$ or $w_3=0$ can be obtained from the 
first or second solutions, \eqref{L5V-1} or \eqref{L5V-2}, respectively,
by reversal of direction of all arrows in Fig.~\ref{fig-SixVertices}, that correspond
to exchange of the weighs $w_1\leftrightarrow w_2$, $w_3\leftrightarrow w_4$, 
$w_5\leftrightarrow w_6$, and hence to the transformation 
\begin{equation}
L\mapsto (\sigma^x\otimes \sigma^x)L(\sigma^x\otimes \sigma^x).
\end{equation}
Correspondingly, the $R$-matrix intertwining these $L$-operators 
is given by the transposed that of \eqref{Rmatrix},  
$(\sigma^x\otimes\sigma^x)R(\sigma^x\otimes\sigma^x)=R^\mathsf{T}$.
The parameter $\Delta$ in the model with $w_1=0$ can be obtained from 
\eqref{Delta-w2} to be 
\begin{equation}
\Delta=\frac{w_3w_4-w_5w_6}{w_2w_4},
\end{equation}
and in the model with $w_3=0$ to be 
\begin{equation}
\Delta=\frac{w_1w_2-w_5w_6}{w_2 w_4}.
\end{equation}

The special case of the finite-dimensional solution given by $L$-operator 
\eqref{ThirdL} at $m=1$ describes the four-vertex model \cite{B-08a,B-08b}. 
Specifically, within our choice this is the model with the weights
\begin{equation}\label{fourw}
w_1=-\frac{1}{\Delta u}, \qquad 
w_3=u,\qquad
w_4=w_2=0,\qquad
w_5=w_6=1.
\end{equation}
In this model 
\begin{equation}\label{four-D}
\Delta=-w_5w_6/w_1w_3, 
\end{equation}
that can be obtained by setting $w_4=0$ in 
\eqref{Delta-w2} or $w_2=0$ in \eqref{Delta-w4}. 

Note that, the four-vertex model described by \eqref{fourw} and \eqref{four-D}
is one out of four possible four-vertex models which can be considered   
in the context of the general six-vertex model. They arise according to 
paring of the weights $w_1$ and $w_2$ with $w_3$ and $w_4$, such that 
$w_i=w_k=0$, $i=1,2$, $k=3,4$. The remaining three possible choices 
can be obtained from  \eqref{fourw} and \eqref{four-D} by reversal of direction 
of all arrows ($w_1\leftrightarrow w_2$, $w_3\leftrightarrow w_4$, 
$w_5\leftrightarrow w_6$) and exchange of assignment of 
the quantum and auxiliary spaces of the $L$-operator 
to the vertical and horizontal lines of the lattice. The latter 
corresponds to the mapping $L\mapsto PLP$,  where $P$ is the permutation operator, 
or exchange $w_3\leftrightarrow w_4$, $w_5\leftrightarrow w_6$ in  \eqref{L6v}.

The four-vertex model is known to be related to 
special enumerations of boxed plane partitions, 
with the fixed sums of their diagonals \cite{BM-19}.

\section{The scalar product case}

In \cite{B-10} (see also \cite{MSa-13,MSa-14}), 
it was shown that the scalar product of 
$n$-particle off-shell Bethe states (for the value $\Delta=1$) 
is given by
\begin{equation}\label{Bogoliubov}
\bra{\Omega}
\prod_{j=1}^{n}C(v_j)
\prod_{j=1}^{n}B(u_j)
\ket{\Omega}
=
\prod_{1\le j<k\le n}
\frac{1}{(v_j^2-v_k^2)(u_k^2-u_j^2)}
\det Q_n,
\end{equation}
where $Q_n$ is the following $n\times n$ matrix
\begin{equation}\label{Qmat}
(Q_n)_{jk}=\frac{d(v_j)a(u_k)u_k^{2(n-1)}
-d(u_k)a(v_j)v_j^{2(n-1)}}
{u_k/v_j-v_j/u_k},
\qquad j,k=1,\ldots,n.
\end{equation}
However, the representation for the scalar product which is provided by
\eqref{ProdA} or \eqref{ProdD} at $m=0$ 
is given in terms of an $2n\times 2n$ determinant of the matrix $\mathcal{V}_{2n,n}$, 
with entries \eqref{VLn}. Here, we show that these two representations, in terms 
of $2n\times 2n$ and $n\times n$ determinants, are equivalent. 

In accordance with \eqref{CB},
we assign $v_j=u_{2n-j+1}$, $j=1,\ldots,n$. For abbreviation, 
we introduce the following notation 
\begin{equation}
\tilde{a}_j=a(u_j)u_j^{2n-1},
\qquad
\tilde{d}_j=d(u_j)u_j.
\end{equation}
The matrix in \eqref{Qmat} then reads 
\begin{equation}\label{Qmat2}
(Q_n)_{ij}=\frac{\tilde{d}_{2n-i+1}\tilde{a}_{j}
-\tilde{d}_{j}\tilde{a}_{2n-i+1}}{u_j^2-u_{2n-i+1}^2}.
\end{equation}
The entries of the matrix $\mathcal{V}_{2n,n}$, see \eqref{VLn}, are given by
\begin{equation}
(\mathcal{V}_{2n,n})_{ij}=
\begin{cases}
\tilde{d}_ju_j^{2(i-1)}&i=1,\ldots,n,
\\[4pt]
\tilde{a}_ju_j^{2(i-n-1)} &i=n+1,\ldots,2n.
\end{cases}
\end{equation}
The equivalence of the two representations for the scalar product implies 
the following identity to hold
\begin{equation}\label{det2n=detn}
\prod_{1\leq i<j\leq L}\frac{1}{u_j^2-u_i^2}
\det\mathcal{V}_{2n,n}
=
\prod_{1\le j<k\le n}
\frac{1}{(u_k^2-u_j^2)(u_{n+k}^2-u_{n+j}^2)}
\det Q_n.
\end{equation}

To prove \eqref{det2n=detn} we rewrite both sides using the definition of the determinant 
as the sum over permutations in the form
\begin{equation}\label{detV2nn=sum}
\prod_{1\leq i<j\leq L}\frac{1}{u_j^2-u_i^2}
\det\mathcal{V}_{2n,n}
=
\sum_{\sigma\in S_{2n}} C_{2n\times 2n}(\sigma)
\prod_{i=1}^{n}\tilde{d}_{\sigma(i)}\tilde{a}_{\sigma(n+i)}
\end{equation}
and
\begin{equation}\label{detQ=sum}
\prod_{1\le j<k\le n}
\frac{1}{(u_k^2-u_j^2)(u_{n+k}^2-u_{n+j}^2)}
\det Q_n
=
\sum_{\sigma\in S_{2n}}C_{n\times n}(\sigma)
\prod_{i=1}^{n}\tilde{d}_{\sigma(i)}\tilde{a}_{\sigma(n+i)},
\end{equation}
respectively, where the summation is performed over elements 
of the symmetric group $S_{2n}$---permutations of $2n$ elements
$\sigma:(1,\ldots,2n)\mapsto(\sigma(1),\ldots,\sigma(2n))$. 
The coefficients $C_{2n\times 2n}(\sigma)$ and $C_{n\times n}(\sigma)$
depend on the variables $u_1,\ldots,u_{2n}$ and one can prove
\eqref{det2n=detn} by showing that $C_{2n\times 2n}(\sigma)=C_{n\times n}(\sigma)$.

The expressions in \eqref{detQ=sum} and \eqref{detV2nn=sum} are apparently 
symmetric with respect to permutations 
of $u_1,\ldots,u_n$ and of $u_{n+1},\ldots,u_{2n}$, therefore, we need to compare
only the coefficients in both expressions of the terms
\begin{equation}
\tilde{d}_1\cdots\tilde{d}_m
\tilde{a}_{m+1}\cdots\tilde{a}_n
\tilde{d}_{n+1}\cdots\tilde{d}_{2n-m}
\tilde{a}_{2n-m+1}\cdots\tilde{a}_{2n}, 
\qquad m=0,1,\ldots,n.
\end{equation}
Let us denote the permutations $\sigma\in S_{2n}$, which corresponds to 
these terms in \eqref{detQ=sum} and \eqref{detV2nn=sum} as $\sigma_m$, that is 
\begin{equation}
\sigma_m:(1,\ldots,2n)\mapsto 
(1,\ldots,m,n+1,\ldots,2n-m,m+1,\ldots,n,2n-m+1,\ldots,2n).
\end{equation}
The coefficients $C_{2n\times 2n}(\sigma_m)$ and $C_{n\times n}(\sigma_m)$
can be computed from \eqref{detQ=sum} and \eqref{detV2nn=sum}, respectively, 
by setting in both sides of these relations
\begin{equation}\label{da=1010}
\begin{split}
\tilde{d}_i 
&\mapsto
\begin{cases}
1 & i=1,\ldots,m,n+1,\ldots,2n-m\\
0 & i=m+1,\ldots,n,2n-m+1,\ldots,2n,
\end{cases}
\\
\tilde{a}_i 
&\mapsto 
\begin{cases}
1& i=m+1,\ldots,n,2n-m+1,\ldots,2n\\
0& i=1,\ldots,m,n+1,\ldots,2n-m.
\end{cases}
\end{split}
\end{equation}

Let us consider $C_{2n\times 2n}(\sigma_m)$. 
The matrix $\mathcal{V}_{2n,n}$ up on setting the values 
of $\tilde{d}$'s and $\tilde{a}$'s according to \eqref{da=1010},  
after moving columns becomes a block Vandermonde matrix. More precisely, we need 
to swap the set of columns from $m$ to $n$ with set of columns from 
$n+1$ to $2n-m$ as a whole sets. This rearranging 
is in fact corresponds to the permutation $\sigma_m$ and give us the factor 
$(-1)^{n-m}=\sgn(\sigma_m)$. Hence,
\begin{multline}
C_{2n\times 2n}(\sigma_m)
=\sgn(\sigma_m) 
\prod_{1\leq i<j\leq 2n}\frac{1}{u_j^2-u_i^2}
\\ \times
\prod_{1\leq i<j\leq n}\big(u_{\sigma_m(j)}^2-u_{\sigma_m(i)}^2\big)
\big(u_{\sigma_m(n+j)}^2-u_{\sigma_m(n+i)}^2\big).
\end{multline}
Now, using that $\prod_{i<j}(u_j^2-u_i^2)=\sgn(\sigma_m)\prod_{i<j}
\big(u_{\sigma_m(j)}^2-u_{\sigma_m(i)}^2\big)$, and simplifying factors, 
we obtain
\begin{equation}\label{C2n2n}
C_{2n\times 2n}(\sigma_m)
=\prod_{i,j=1}^n\frac{1}{u_{\sigma_m(n+j)}^2-u_{\sigma_m(i)}^2}.
\end{equation}

Let us calculate $C_{n\times n}(\sigma_m)$. Using \eqref{da=1010}, from  
\eqref{Qmat2} and \eqref{detQ=sum} we get
\begin{equation}\label{Cnn}
C_{n\times n}(\sigma_m)=
\prod_{1\le i<j\le n}
\frac{1}{(u_j^2-u_i^2)(u_{n+j}^2-u_{n+i}^2)}
\det A_{m} \det D_{n-m},
\end{equation}
where $A_{m}$ and $D_{n-m}$ 
are $m\times m$ and $(n-m)\times (n-m)$ matrices, respectively, with entries
\begin{equation}
\left(A_{m}\right)_{ij}=
\frac{1}{u_{2n-i+1}^2-u_j^2},
\quad
\left(D_{n-m}\right)_{ij}=
\frac{1}{u_{m+j}^2-u_{2n-i-m+1}^2}.
\end{equation}
The matrices $A_{m}$ and $D_{n-m}$ are Cauchy matrices,
and a general formula for the determinant of such matrix is 
\begin{equation}\label{Cauchy-det}
\det_{1\leq i,j\leq n} 
\left[\frac{1}{x_i-y_j}\right]=
\prod_{i,j=1}^{n} \frac{1}{x_i-y_j}
\prod_{1\leq i<j\leq n} (x_j-x_i)(y_i-y_j).
\end{equation}
Using \eqref{Cauchy-det}, we obtain
\begin{equation}\label{detAm}
\det A_m = 
\prod_{i,j=1}^{m}
\frac{1}{u_{2n-m+j}^2-u_j^2}
\prod_{1\leq i<j\leq m}
\left(u_j^2-u_i^2\right)\left(u_{2n-m+j}^2-u_{2n-m+i}^2\right)
\end{equation}
and
\begin{multline}\label{detDnm}
\det D_{n-m} =(-1)^{n-m}
\prod_{i,j=1}^{n-m}
\frac{1}{u_{m+j}^2-u_{n+i}^2}
\\ \times
\prod_{1\leq i<j\leq n-m}
\left(u_{m+j}^2-u_{m+i}^2\right)\left(u_{n+j}^2-u_{n+i}^2\right).
\end{multline}
Now, rewriting the double products in \eqref{Cnn} as
\begin{multline}
\prod_{1\leq i<j\leq n}^{}\left(u_j^2-u_i^2\right)
= \prod_{1\leq i<j\leq m}^{}\left(u_j^2-u_i^2\right)
\\ \times
\prod_{1\leq i<j\leq n-m}^{}\left(u_{m+j}^2-u_{m+i}^2\right)
\prod_{j=1}^{m}
\prod_{i=1}^{n-m}\left(u_{m+i}^2-u_j^2\right)
\end{multline}
and 
\begin{multline}
\prod_{1\leq i<j\leq n}^{}\left(u_{n+j}^2-u_{n+i}^2\right)
= \prod_{1\leq i<j\leq n-m}^{}\left(u_{n+j}^2-u_{n+i}^2\right)
\\ \times
\prod_{1\leq i<j\leq m}^{}\left(u_{2n-m+j}^2-u_{2n-m+i}^2\right)
\prod_{j=1}^{m}
\prod_{i=1}^{n-m}\left(u_{2n-m+j}^2-u_{n+i}^2\right),
\end{multline}
and comparing these expressions with 
those in \eqref{detAm} and \eqref{detDnm}, we see that 
all the factors in the numerator in \eqref{Cnn} 
simplifies, so that we leave with the expression
\begin{multline}
C_{n\times n}(\sigma_m)= 
(-1)^{n-m} \prod_{i,j=1}^{m}\frac{1}{u_{2n-m+i}^2-u_j^2}
\prod_{i,j=1}^{n-m}
\frac{1}{u_{m+j}^2-u_{n+i}^2}
\\ \times
\prod_{j=1}^{m}
\prod_{i=1}^{n-m}\frac{1}{\left(u_{m+i}^2-u_j^2\right)
\left(u_{2n-m+j}^2-u_{n+i}^2\right)}.
\end{multline}
But this expression is exactly that standing in \eqref{C2n2n}.  
Thus we have just showed that representation \eqref{Bogoliubov}
is equivalent to \eqref{ProdA} or \eqref{ProdD} in the scalar product 
case.

As a final comment here, let us note that the coefficients 
$C_{2n\times 2n}(\sigma)$ and $C_{n\times n}(\sigma)$
can be written for an arbitrary permutation $\sigma\in S_{2n}$
in the form
\begin{equation}
C_{2n\times 2n}(\sigma)=C_{n\times n}(\sigma)
=\prod_{i,j=1}^n\frac{1}{u_{\sigma(n+j)}^2-u_{\sigma(i)}^2}, 
\end{equation}
in agreement with the total symmetry of 
the right-hand sides of \eqref{detQ=sum} and \eqref{detV2nn=sum}
with respect to permutations of the rapidity variables.

\bibliography{fvdets_bib.bib}
\end{document}

%% file: fig-SixVertices.tex
\begin{tikzpicture}[scale=.5]
\draw [semithick] (0.2,3)--(1.8,3);
\draw [semithick] (1,2.2)--(1,3.8);
\draw [thick] [->] (0.5,3)--(.6,3);
\draw [thick] [->] (1.5,3)--(1.6,3);
\draw [thick] [->] (1,2.5)--(1,2.6);
\draw [thick] [->] (1,3.5)--(1,3.6);
\draw [semithick] (0.2,1)--(1.8,1);
\draw [semithick] (1,0.2)--(1,1.8);
\node at (1,-1) {$w_1$};
\end{tikzpicture}
\quad
\begin{tikzpicture}[scale=.5]
\draw [semithick] (0.2,3)--(1.8,3);
\draw [semithick] (1,2.2)--(1,3.8);
\draw [thick] [->] (0.5,3)--(0.4,3);
\draw [thick] [->] (1.5,3)--(1.4,3);
\draw [thick] [->] (1,2.5)--(1,2.4);
\draw [thick] [->] (1,3.5)--(1,3.4);
\draw [line width=3] (0.2,1)--(1.8,1);
\draw [line width=3] (1,0.2)--(1,1.8);
\node at (1,-1) {$w_2=0$};
\end{tikzpicture}
\quad
\begin{tikzpicture}[scale=.5]
\draw [semithick] (0.2,3)--(1.8,3);
\draw [semithick] (1,2.2)--(1,3.8);
\draw [thick] [->] (0.5,3)--(0.6,3);
\draw [thick] [->] (1.5,3)--(1.6,3);
\draw [thick] [->] (1,2.5)--(1,2.4);
\draw [thick] [->] (1,3.5)--(1,3.4);
\draw [semithick] (0.2,1)--(1.8,1);
\draw [line width=3] (1,0.2)--(1,1.8);
\node at (1,-1) {$w_3$};
\end{tikzpicture}
\quad
\begin{tikzpicture}[scale=.5]
\draw [semithick] (0.2,3)--(1.8,3);
\draw [semithick] (1,2.2)--(1,3.8);
\draw [thick] [->] (0.5,3)--(0.4,3);
\draw [thick] [->] (1.5,3)--(1.4,3);
\draw [thick] [->] (1,2.5)--(1,2.6);
\draw [thick] [->] (1,3.5)--(1,3.6);
\draw [line width=3] (0.2,1)--(1.8,1);
\draw [semithick] (1,0.2)--(1,1.8);
\node at (1,-1) {$w_4$};
\end{tikzpicture}
\quad
\begin{tikzpicture}[scale=.5]
\draw [semithick] (0.2,3)--(1.8,3);
\draw [semithick] (1,2.2)--(1,3.8);
\draw [thick] [->] (0.5,3)--(.6,3);
\draw [thick] [->] (1.5,3)--(1.4,3);
\draw [thick] [->] (1,2.5)--(1,2.4);
\draw [thick] [->] (1,3.5)--(1,3.6);
\draw [semithick] (0.2,1)--(1,1)--(1,1.8);
\draw [line width=3] (1,0.2)--(1,1)--(1.8,1);
\node at (1,-1) {$w_5$};
\end{tikzpicture}
\quad
\begin{tikzpicture}[scale=.5]
\draw [semithick] (0.2,3)--(1.8,3);
\draw [semithick] (1,2.2)--(1,3.8);
\draw [thick] [->] (0.5,3)--(.4,3);
\draw [thick] [->] (1.5,3)--(1.6,3);
\draw [thick] [->] (1,2.5)--(1,2.6);
\draw [thick] [->] (1,3.5)--(1,3.4);
\draw [line width=3] (0.2,1)--(1,1)--(1,1.8);
\draw [semithick] (1,0.2)--(1,1)--(1.8,1);
\node at (1,-1) {$w_6$};
\end{tikzpicture}

%% file: fig-LxMlattice.tex

\usetikzlibrary{decorations.pathreplacing}

\begin{tikzpicture}[scale=.45]
\draw [semithick] (0.2,1)--(8.8,1);
\draw [semithick] (0.2,2)--(8.8,2);
\draw [semithick] (0.2,3)--(8.8,3);
\draw [semithick] (0.2,4)--(8.8,4);
\draw [semithick] (0.2,5)--(8.8,5);
\draw [semithick] (0.2,6)--(8.8,6);
\draw [semithick] (0.2,7)--(8.8,7);
\draw [semithick] (0.2,8)--(8.8,8);
\draw [semithick] (0.2,9)--(8.8,9);
\draw [semithick] (1,0.2)--(1,9.8);
\draw [semithick] (2,0.2)--(2,9.8);
\draw [semithick] (3,0.2)--(3,9.8);
\draw [semithick] (4,0.2)--(4,9.8);
\draw [semithick] (5,0.2)--(5,9.8);
\draw [semithick] (6,0.2)--(6,9.8);
\draw [semithick] (7,0.2)--(7,9.8);
\draw [semithick] (8,0.2)--(8,9.8);
\draw [thick] [->] (.5,1)--(.6,1);
\draw [thick] [->] (.5,2)--(.6,2);
\draw [thick] [->] (.5,3)--(.6,3);
\draw [thick] [->] (.5,4)--(.6,4);
\draw [thick] [->] (.5,5)--(.6,5);
\draw [thick] [->] (.5,6)--(.6,6);
\draw [thick] [->] (.5,7)--(.6,7);
\draw [thick] [->] (.5,8)--(.6,8);
\draw [thick] [->] (.5,9)--(.6,9);
\draw [thick] [->] (1,9.5)--(1,9.6);
\draw [thick] [->] (2,9.5)--(2,9.6);
\draw [thick] [->] (3,9.5)--(3,9.6);
\draw [thick] [->] (4,9.5)--(4,9.6);
\draw [thick] [->] (5,9.5)--(5,9.6);
\draw [thick] [->] (6,9.5)--(6,9.4);
\draw [thick] [->] (7,9.5)--(7,9.4);
\draw [thick] [->] (8,9.5)--(8,9.4);
\draw [thick] [->] (8.5,1)--(8.6,1);
\draw [thick] [->] (8.5,2)--(8.6,2);
\draw [thick] [->] (8.5,3)--(8.6,3);
\draw [thick] [->] (8.5,4)--(8.6,4);
\draw [thick] [->] (8.5,5)--(8.6,5);
\draw [thick] [->] (8.5,6)--(8.6,6);
\draw [thick] [->] (8.5,7)--(8.6,7);
\draw [thick] [->] (8.5,8)--(8.6,8);
\draw [thick] [->] (8.5,9)--(8.6,9);
\draw [thick] [->] (1,.5)--(1,.4);
\draw [thick] [->] (2,.5)--(2,.4);
\draw [thick] [->] (3,.5)--(3,.4);
\draw [thick] [->] (4,.5)--(4,.6);
\draw [thick] [->] (5,.5)--(5,.6);
\draw [thick] [->] (6,.5)--(6,.6);
\draw [thick] [->] (7,.5)--(7,.6);
\draw [thick] [->] (8,.5)--(8,.6);
\draw [decorate,decoration={brace}]
(5.8,10.1) -- (8.2,10.1) node [midway,yshift=9pt] {$N$};
\draw [decorate,decoration={brace}]
(-.1,0.8) -- (-.1,9.2) node [midway,xshift=-9pt] {$M$};
\draw [decorate,decoration={brace}]
(0.8,11.4) -- (8.2,11.4) node [midway,yshift=9pt] {$L$};
\draw [decorate,decoration={brace,mirror}]
(.8,-.1) -- (3.2,-.1) node [midway,yshift=-10pt] {$N$};
\node at (4.5,-2) {(a)};
\end{tikzpicture}
\qquad
\begin{tikzpicture}[scale=.45]
\draw [semithick] (0.2,1)--(8.8,1);
\draw [semithick] (0.2,2)--(8.8,2);
\draw [semithick] (0.2,3)--(8.8,3);
\draw [semithick] (0.2,4)--(8.8,4);
\draw [semithick] (0.2,5)--(8.8,5);
\draw [semithick] (0.2,6)--(8.8,6);
\draw [semithick] (0.2,7)--(8.8,7);
\draw [semithick] (0.2,8)--(8.8,8);
\draw [semithick] (0.2,9)--(8.8,9);
\draw [semithick] (1,0.2)--(1,9.8);
\draw [semithick] (2,0.2)--(2,9.8);
\draw [semithick] (3,0.2)--(3,9.8);
\draw [semithick] (4,0.2)--(4,9.8);
\draw [semithick] (5,0.2)--(5,9.8);
\draw [semithick] (6,0.2)--(6,9.8);
\draw [semithick] (7,0.2)--(7,9.8);
\draw [semithick] (8,0.2)--(8,9.8);
\draw [line width=3] (1,0.2)--(1,4)--(3,4)--(3,5)--(4,5)--(4,6)--(5,6)--(5,9)--(6,9)--(6,9.8);
\draw [line width=3] (2,0.2)--(2,3)--(5,3)--(5,5)--(6,5)--(6,7)--(7,7)--(7,9.8);
\draw [line width=3] (3,0.2)--(3,1)--(6,1)--(6,3)--(8,3)--(8,9.8);
\node at (4.5,-2) {(b)};
\end{tikzpicture}

%% file: fig-3DYoung.tex


\begin{tikzpicture}[scale=.5]
\draw [line width=3] (1,3)--(1,4)--(3,4)--(3,5)--(4,5)--(4,6)--(5,6)--(5,9)--(6,9)--(6,9);
\draw [line width=3] (2,2)--(2,3)--(5,3)--(5,5)--(6,5)--(6,7)--(7,7)--(7,8);
\draw [line width=3] (3,1)--(3,1)--(6,1)--(6,3)--(8,3)--(8,7);
\draw [semithick] (.5,.5)--(.5,9.5)--(8.5,9.5)--(8.5,.5)--(.5,.5);
\draw [semithick] (.5,3.5)--(3.5,.5);
\draw [semithick] (.5,4.5)--(2.5,2.5);
\draw [semithick] (1.5,4.5)--(3.5,2.5);
\draw [semithick] (2.5,4.5)--(4.5,2.5);
\draw [semithick] (2.5,5.5)--(3.5,4.5);
\draw [semithick] (3.5,5.5)--(6.5,2.5);
\draw [semithick] (3.5,6.5)--(5.5,4.5);
\draw [semithick] (4.5,6.5)--(6.5,4.5);
\draw [semithick] (4.5,7.5)--(6.5,5.5);
\draw [semithick] (4.5,8.5)--(6.5,6.5);
\draw [semithick] (4.5,9.5)--(5.5,8.5);
\draw [semithick] (5.5,9.5)--(8.5,6.5);
\draw [semithick] (6.5,7.5)--(8.5,5.5);
\draw [semithick] (7.5,5.5)--(8.5,4.5);
\draw [semithick] (7.5,4.5)--(8.5,3.5);
\draw [semithick] (7.5,3.5)--(8.5,2.5);
\draw [semithick] (6.5,3.5)--(7.5,2.5);
\draw [semithick] (4.5,3.5)--(6.5,1.5);
\draw [semithick] (5.5,1.5)--(6.5,0.5);
\draw [semithick] (4.5,1.5)--(5.5,0.5);
\draw [semithick] (3.5,1.5)--(4.5,0.5);
\draw [semithick] (1.5,9.5)--(1.5,4.5);
\draw [semithick] (2.5,9.5)--(2.5,4.5);
\draw [semithick] (3.5,9.5)--(3.5,5.5);
\draw [semithick] (4.5,9.5)--(4.5,6.5);
\draw [semithick] (5.5,8.5)--(5.5,5.5);
\draw [semithick] (6.5,8.5)--(6.5,7.5);
\draw [semithick] (3.5,4.5)--(3.5,3.5);
\draw [semithick] (4.5,5.5)--(4.5,3.5);
\draw [semithick] (5.5,4.5)--(5.5,1.5);
\draw [semithick] (6.5,6.5)--(6.5,3.5);
\draw [semithick] (7.5,7.5)--(7.5,3.5);
\draw [semithick] (7.5,2.5)--(7.5,0.5);
\draw [semithick] (6.5,2.5)--(6.5,0.5);
\draw [semithick] (4.5,2.5)--(4.5,1.5);
\draw [semithick] (3.5,2.5)--(3.5,1.5);
\draw [semithick] (2.5,2.5)--(2.5,1.5);
\draw [semithick] (1.5,3.5)--(1.5,2.5);
\draw [semithick] (0.5,8.5)--(4.5,8.5);
\draw [semithick] (5.5,8.5)--(6.5,8.5);
\draw [semithick] (0.5,7.5)--(4.5,7.5);
\draw [semithick] (5.5,7.5)--(6.5,7.5);
\draw [semithick] (0.5,6.5)--(4.5,6.5);
\draw [semithick] (6.5,6.5)--(7.5,6.5);
\draw [semithick] (0.5,5.5)--(3.5,5.5);
\draw [semithick] (6.5,5.5)--(7.5,5.5);
\draw [semithick] (0.5,4.5)--(2.5,4.5);
\draw [semithick] (3.5,4.5)--(4.5,4.5);
\draw [semithick] (5.5,4.5)--(7.5,4.5);
\draw [semithick] (1.5,3.5)--(4.5,3.5);
\draw [semithick] (5.5,3.5)--(7.5,3.5);
\draw [semithick] (2.5,2.5)--(5.5,2.5);
\draw [semithick] (6.5,2.5)--(8.5,2.5);
\draw [semithick] (2.5,1.5)--(5.5,1.5);
\draw [semithick] (6.5,1.5)--(8.5,1.5);
\draw [->] (0.5,3.5)--(-1.5,3.5);
\draw [->] (8.5,0.5)--(9.5,-.5);
\draw [->] (5.5,9.5)--(5.5,11);
\node at (-1,3) {$x$};
\node at (8.75,-.5) {$y$};
\node at (5,10.5) {$z$};
\node at (0,4) {$A$};
\node at (9,.75) {$B$};
\node at (6,10) {$C$};
\end{tikzpicture}
\qquad 
\begin{tikzpicture}[scale=.5]
\draw [semithick] (0.5,0.5)--(0.5,1.5)--(1.5,1.5)--(1.5,.5)--(.5,.5);
\draw [semithick] (0.2,3)--(1.8,3);
\draw [semithick] (1,2.2)--(1,3.8);
\node at (1,-1) {};
\end{tikzpicture}
\quad
\begin{tikzpicture}[scale=.5]
\draw [semithick] (0.5,0.5)--(0.5,1.5)--(1.5,.5)--(1.5,1.5);
\draw [line width=3] (1,.5)--(1,1.5);
\draw [semithick] (.2,3)--(1.8,3);
\draw [line width=3] (1,2.2)--(1,3.8);
\node at (1,-1) {};
\end{tikzpicture}
\quad
\begin{tikzpicture}[scale=.5]
\draw [semithick] (.5,.5)--(1.5,.5)--(.5,1.5)--(1.5,1.5);
\draw [line width=3] (.5,1)--(1.5,1);
\draw [line width=3] (.2,3)--(1.8,3);
\draw [semithick] (1,2.2)--(1,3.8);
\node at (1,-1) {};
\end{tikzpicture}
\quad
\begin{tikzpicture}[scale=.5]
\draw [semithick] (.5,.5)--(.5,1.5)--(1.5,1.5);
\draw [semithick] (1.5,.5)--(.5,1.5);
\draw [line width=3] (1,.5)--(1,1)--(1.5,1);
\draw [semithick] (.2,3)--(1,3)--(1,3.8);
\draw [line width=3] (1,2.2)--(1,3)--(1.8,3);
\node at (1,-1) {};
\end{tikzpicture}
\quad
\begin{tikzpicture}[scale=.5]
\draw [semithick] (.5,.5)--(1.5,.5)--(1.5,1.5);
\draw [semithick] (1.5,.5)--(.5,1.5);
\draw [line width=3] (.5,1)--(1,1)--(1,1.5);
\draw [line width=3] (.2,3)--(1,3)--(1,3.8);
\draw [semithick] (1,2.2)--(1,3)--(1.8,3);
\node at (1,-1) {};
\end{tikzpicture}


%% file: fig-Lop.tex

\begin{tikzpicture}[scale=.5]
\draw [semithick] (0,2)--(3.6,2)--(3.6,2.1)--(4,2)--(3.6,1.9)--(3.6,2);
\draw [semithick] (2,0)--(2,3.6)--(1.9,3.6)--(2,4)--(2.1,3.6)--(2,3.6);
\node at (2,4.75) {$u_\mu$};
\node at (4.75,2) {$\xi_k$};
\end{tikzpicture}

%% file: fig-ABCD.tex

\begin{tikzpicture}[scale = .5]
\draw [semithick] (1,5)--(1,0) node [below] {$A(u)$};
\draw [semithick] (0,1)--(2,1);
\draw [semithick] (0,2)--(2,2);
\draw [semithick] (0,3)--(2,3);
\draw [semithick] (0,4)--(2,4);
\draw [thick] [->] (1,0.5)--(1,0.6);
\draw [thick] [->] (1,4.5)--(1,4.6);
\node at (1,5.5) {$u$};
\end{tikzpicture}
\quad
\begin{tikzpicture}[scale = .5]
\draw [semithick] (1,5)--(1,0) node [below] {$B(u)$};
\draw [semithick] (0,1)--(2,1);
\draw [semithick] (0,2)--(2,2);
\draw [semithick] (0,3)--(2,3);
\draw [semithick] (0,4)--(2,4);
\draw [thick] [->] (1,0.5)--(1,0.6);
\draw [thick] [->] (1,4.6)--(1,4.5);
\node at (1,5.5) {$u$};
\end{tikzpicture}
\quad
\begin{tikzpicture}[scale = .5]
\draw [semithick] (1,5)--(1,0) node [below] {$C(u)$};
\draw [semithick] (0,1)--(2,1);
\draw [semithick] (0,2)--(2,2);
\draw [semithick] (0,3)--(2,3);
\draw [semithick] (0,4)--(2,4);
\draw [thick] [->] (1,0.6)--(1,0.5);
\draw [thick] [->] (1,4.5)--(1,4.6);
\node at (1,5.5) {$u$};
\end{tikzpicture}
\quad
\begin{tikzpicture}[scale = .5]
\draw [semithick] (1,5)--(1,0) node [below] {$D(u)$};
\draw [semithick] (0,1)--(2,1);
\draw [semithick] (0,2)--(2,2);
\draw [semithick] (0,3)--(2,3);
\draw [semithick] (0,4)--(2,4);
\draw [thick] [->] (1,0.6)--(1,0.5);
\draw [thick] [->] (1,4.6)--(1,4.5);
\node at (1,5.5) {$u$};
\node at (3,4) {$\xi_1$};
\node at (3,1) {$\xi_M$};
\node at (2.9,2.7) {$\vdots$};
\end{tikzpicture}
